\documentstyle[aps]{revtex}

\begin{document}
\author{Yong-Ping Zhang, Hui Yu, Ying-Fang Gao, J. -Q. Liang}
\address{Institute of Theoretical Physics, Shanxi\\
University, Taiyuan, Shanxi, 030006, China}
\title{Quantum transport through a double Aharonov-Bohm-interferometer in
the presence of Andreev reflection}
\maketitle

\begin{abstract}
Quantum transport through a double Aharonov-Bohm-interferometer in the
presence of Andreev reflection is investigated in terms of the
nonequilibrium Green function method with which the reflection current is
obtained. Tunable Andreev reflection probabilities depending on the interdot
coupling strength and magnetic flux as well are analysised in detail. It is
found that the oscillation period of the reflection probability with respect
to the magnetic flux for the double interferometer depends linearly on the
ratio of two parts{\bf \ }magnetic{\bf \ }fluxes{\bf \ }$n${\bf ,} i.e. $%
2(n+1)\pi ${\bf , }while that of a single interferometer is{\bf \ }$2\pi $%
{\bf .} The coupling strength not only affects the height and the linewidth
of Andreev reflection current peaks vs gate votage but also shifts the peak
positions. It is furthermore demonstrated that the Andreev reflection
current peaks can be tuned by the magnetic fluxes.
\end{abstract}


\section{ INTRODUCTION}

Quantum dots (QDs) containing discrete energy levels which can be tuned
easily in experiment have been attracting considerable attention
continuously, and the study of quantum transport properties through QDs has
become an interesting field which serves as a testing ground for the
physical phenomena, such as Kondo \cite{1,2} and Fano effects\cite{3}.
Recent developments on the study include Coulomb on-site repulsion \cite{4},
spin-flip effect caused by the magnetic leads \cite{5} or rotating magnetic
field \cite{6}, and phonon-assisted tunneling in the molecular QD \cite{7}%
with time-dependent field or ac gate voltage \cite{8,9}.

Electron transports coherently through an open QD can be investigated in the
standard procedure by means of nonequilibrium Green function (NGF) \cite{8},
the rate equation \cite{10} and some other methods\cite{11}. The current
characteristic properties change dramatically if the QD is connected to
additional devices, so various kinds of hybrid QD systems have been studied
in the literature. Two aspects mainly involved in those hybrid systems are
the ferromagnetic and superconductor properties which evoke the spin degree
of freedom of electrons. The F-QD-F system, i.e., QD coupled with two
ferromagnetic leads, has been studied as spin-valve device \cite{5,12}. In
the superconductor case, the Andreev reflection, conserving the momentum of
the reflected particle but not the charge, plays a dominant role at the
interface between a normal metal and a superconductor. Using multichannel
S-matrix method, Beenakker demonstrated the resonant Andreev tunneling in
the normal metal-normal QD-superconductor (N-NQD-S) hybrid system in the
zero-bias limit \cite{13}; later Claughton et al. extended this theory to
the finite-bias case \cite{14}. Sun et al. explored the resonant Andreev
reflection in this system with multienergy levels in normal QD and found
that some extra peaks originating from the Andreev reflection superimposed
on the conventional current plateaus \cite{15}. And they also investigated
the photon-assisted Andreev and normal tunnels in this system. With an
applied time-varying external field, it was observed that a series of
photon-assisted Andreev tunneling peaks appeared in the curve of the average
current $\left\langle I\right\rangle $ vs the gate voltage with the negative
peaks on the left-hand side and positive peaks on the right-hand side of the
original resonant peak in the absence of the external field \cite{16}.
Another interesting hybrid system consists of ferromagnet and superconductor
leads coupled with a normal QD (F-QD-S). In Ref. \cite{17}, Zhu et al.
provided an efficient method for introducing a spin degree of freedom in
this system, and also derived the general current formula of the Andreev
reflection in a quantum dot connected with two ferromagnetic and one
superconductor leads (F$_{\text{1}}$, F$_{\text{2}}$-QD-S) \cite{18}.
Considering both the Coulomb interaction and the multienergy levels in QD,
Feng and Xiong studied the transport properties in the F-QD-S system \cite%
{19}. Very recently, Cao et al. considered the spin-flip effect in QD of a
F-QD-S system and found that the single-resonance peak of the Andreev
reflection conductance versus the gate voltage would spill into a
double-resonance peak with the increasing of the spin-flip scattering
strength \cite{20}. Thus far, most of studies relating to Andreev reflection
have been focused on a single-QD system.

On the other hand, the electron transport through a double-quantum-dot
(DQD), especially the parallel-coupled double-quantum-dot (PDQD), is an
interesting subject which may involve the quantum phase interference. An
open PDQD threaded by a magnetic flux is regarded as an Aharonov-Bohm (AB)
interferometer in which the coherence of the electron remains partly even in
the presence of Coulomb repulsion\cite{1,21}. Tunable coupling strength
between two parallel dots makes the transport properties more complex and
interesting\cite{22,23}. The tunable Andreev reflection in an AB
interferometer has been investigated recently by Peng et al.\cite{25}.

In this paper we shall study the electron transport through a N-PDQD-S
structure, i.e., a parallel-coupled double-quantum dot that is connected
with two leads, one is a normal metal and the other is a superconductor. The
device threaded by two magnetic fluxes (see Fig.1) with the interdot
coupling can be regarded as a double AB interferometer. By using the NGF
method, the current $J_{L}$ and the probability of the Andreev reflection
are derived.

The paper is organized as follows: In Sec. II, we present the model
Hamiltonian and derive the general formula of the current and the
probability of the Andreev reflection with NGF method. In Sec. III, the
properties of the current due to Andreev reflection depending on bias
voltage, magnetic flux, gate voltage, and interdot coupling strength are
studied in detail, and several dufferent effects are predicted. Finally, a
brief summary is given in Sec. IV.

\section{MODEL HAMILTONIAN AND FORMULATION}

The Hamiltonian of the system schematically shown in Fig. 1 can be
decomposed into four parts of the following form: 
\begin{equation}
H=H_{L}+H_{R}+H_{D}+H_{T},  \eqnum{1}
\end{equation}%
where 
\begin{eqnarray}
H_{L} &=&\sum_{k\sigma }(\varepsilon _{k}-eV)a_{k\sigma }^{\dagger
}a_{k\sigma },  \nonumber \\
H_{R} &=&\sum_{p\sigma }\varepsilon _{p}b_{p\sigma }^{\dagger }b_{p\sigma
}+\sum_{p}(\Delta b_{p\downarrow }b_{-p\uparrow }+h.c.),  \nonumber \\
H_{D} &=&\sum_{\sigma ,i=1,2}\varepsilon _{i}d_{i\sigma }^{\dagger
}d_{i\sigma }+\sum_{\sigma }(\Omega e^{i\Phi _{12}}d_{2\sigma }^{\dagger
}d_{1\sigma }+h.c.),  \eqnum{2} \\
H_{T} &=&\sum_{k\sigma ,i=1,2}(L_{i}e^{i\Phi _{iL}}a_{k\sigma }^{\dagger
}d_{i\sigma }+h.c.)  \nonumber \\
&&+\sum_{p\sigma ,i=1,2}(R_{i}e^{i\Phi _{iR}}b_{p\sigma }^{\dagger
}d_{i\sigma }+h.c.).  \nonumber
\end{eqnarray}%
$H_{L}$ describes the left normal metal lead, $a_{k\sigma }$ ($a_{k\sigma
}^{\dagger }$) is the corresponding annihilation (creation) operator, and $V$
is the bias voltage. We assume the chemical potential of right lead $\mu
_{R}=0$ \cite{16}. $H_{R}$ describes the right superconductor lead with the
energy gap $\Delta $. $H_{D}$ models the coupled double quantum dots, $%
\Omega $ is the interdot coupling strength, and $\Phi _{12}$ is the
corresponding phase shift induced by the magnetic flux. Here we consider the
single energy level in each quantum dot. $H_{T}$ is the tunneling part of
Hamiltonian, $L_{i}$ ($R_{i}$), independent of momentum $k$ ($p$), is the
hopping strength between $i$th quantum dot [i. e. the quantum dot with
energy level $\varepsilon _{i}$ ($i=1,2$)] and left (right) lead, and $\Phi
_{L1}$ is the phase shift of the electron tunneling from the left lead to
first quantum dot, where{\bf \ } 
\[
\Phi _{L1}=\Phi _{1R}=\frac{\Phi }{4} 
\]%
where $\Phi =2\pi (\alpha _{1}+\alpha _{2})${\bf \ }with{\bf \ }$\Phi
_{i}=\alpha _{i}\Phi _{0}${\bf \ (}$i=1,2${\bf ) }being the magnetic flux in
the quantum unit{\bf \ }$\Phi _{0}=hc/e${\bf \ }and{\bf \ }$\Phi _{12}=-\Phi
_{21}=2\pi (\alpha _{1}-\alpha _{2})${\bf \ .}

The current from the left lead to a quantum dot can be written as \cite{4,8}

\begin{eqnarray}
J_{L}(t) &=&-\frac{2e}{\hbar }%
\mathop{\rm Im}%
\int d\varepsilon \sum_{i,j=1,2}\sum_{k\sigma }L_{i}L_{j}^{\ast }e^{i(\Phi
_{iL}+\Phi _{Lj})}  \nonumber \\
&&\times \delta (\varepsilon -\varepsilon _{k})\int_{-\infty }^{\infty
}dt_{1}[G_{ij\sigma }^{r}(t,t_{1})f_{L}(\varepsilon +eV)  \nonumber \\
&&+\theta (t^{\prime }-t_{1})G_{ij\sigma }^{<}(t,t_{1})]e^{-i\varepsilon
(t_{1}-t^{\prime })},  \eqnum{3}
\end{eqnarray}%
where we define $G_{ij\sigma }^{r}(t,t^{\prime })=-i\theta (t-t^{\prime
})\left\langle \left\{ d_{i\sigma }(t),d_{j\sigma }^{\dagger }(t^{\prime
})\right\} \right\rangle $ and $G_{ij\sigma }^{<}(t,t^{\prime
})=i\left\langle d_{j\sigma }^{\dagger }(t^{\prime })d_{i\sigma
}(t)\right\rangle $. Because of the right superconductor lead, it is
convenient to introduce 4$\times $4 matrix representation in which ${\bf G}%
^{r}(t,t^{\prime })$ and ${\bf G}^{<}(t,t^{\prime })$ have the forms,

\begin{equation}
{\bf G}^r(t,t^{\prime })\equiv -i\theta (t-t^{\prime })\left\langle \left\{
\Psi (t),\Psi ^{\dagger }(t^{\prime })\right\} \right\rangle ,  \eqnum{4}
\end{equation}

\begin{equation}
{\bf G}^{<}(t,t^{\prime })\equiv i\left\langle \Psi ^{\dagger }(t^{\prime
})\Psi (t)\right\rangle ,  \eqnum{5}
\end{equation}%
where $\Psi ^{\dagger }$ $=$ $(d_{1\uparrow }^{\dagger },d_{1\downarrow
},d_{2\uparrow }^{\dagger },d_{2\downarrow })$. So Eq. (3) can be rewritten
as 
\begin{equation}
J_{L\uparrow }(t)=-\frac{2e}{\hbar }%
\mathop{\rm Im}%
\int \frac{d\varepsilon }{2\pi }\int_{-\infty }^{t}dt^{\prime
}e^{-i\varepsilon (t-t^{\prime })}\{{\bf \Gamma }^{L}[{\bf G}%
^{r}(t,t^{\prime })f_{L}(\varepsilon +eV)+{\bf G}^{<}(t,t^{\prime
})]\}_{11+33},  \eqnum{6}
\end{equation}%
where 
\begin{eqnarray}
{\bf \Gamma }^{L} &=&2\pi \sum_{k}\delta (\varepsilon -\varepsilon
_{k})\left( 
\begin{array}{cccc}
L_{1}L_{1}^{\ast } & 0 & L_{2}L_{1}^{\ast }e^{i\frac{\Phi }{2}} & 0 \\ 
0 & L_{1}L_{1}^{\ast } & 0 & L_{1}L_{2}^{\ast }e^{-i\frac{\Phi }{2}} \\ 
L_{1}L_{2}^{\ast }e^{-i\frac{\Phi }{2}} & 0 & L_{2}L_{2}^{\ast } & 0 \\ 
0 & L_{2}L_{1}^{\ast }e^{i\frac{\Phi }{2}} & 0 & L_{2}L_{2}^{\ast }%
\end{array}%
\right)  \nonumber \\
&=&\left( 
\begin{array}{cccc}
\Gamma _{1}^{L} & 0 & \sqrt{\Gamma _{1}^{L}\Gamma _{2}^{L}}e^{i\frac{\Phi }{2%
}} & 0 \\ 
0 & \Gamma _{1}^{L} & 0 & \sqrt{\Gamma _{1}^{L}\Gamma _{2}^{L}}e^{-i\frac{%
\Phi }{2}} \\ 
\sqrt{\Gamma _{1}^{L}\Gamma _{2}^{L}}e^{-i\frac{\Phi }{2}} & 0 & \Gamma
_{2}^{L} & 0 \\ 
0 & \sqrt{\Gamma _{1}^{L}\Gamma _{2}^{L}}e^{i\frac{\Phi }{2}} & 0 & \Gamma
_{2}^{L}%
\end{array}%
\right) .  \eqnum{7}
\end{eqnarray}%
with $\Gamma _{1}^{L}=2\pi \sum_{k}\delta (\varepsilon -\varepsilon
_{k})L_{1}L_{1}^{\ast }$, $\Gamma _{2}^{L}=2\pi \sum_{k}\delta (\varepsilon
-\varepsilon _{k})L_{2}L_{2}^{\ast }$, here we consider only a spin-up
current $J_{L\uparrow }(t)$ [ $J_{L\downarrow }(t)$ is easily obtained from $%
J_{L\uparrow }(t)$ by exchanging the spin index].

To calculate ${\bf G}^{r}(t,t^{\prime })$, we start from the Dyson equation, 
\begin{equation}
{\bf G}^{r}(t,t^{\prime })={\bf g}^{r}(t,t^{\prime })+\int dt_{1}dt_{2}{\bf G%
}^{r}(t,t_{1}){\bf \Sigma }^{r}(t_{1},t_{2}){\bf g}^{r}(t_{2},t^{\prime }), 
\eqnum{8}
\end{equation}%
in which ${\bf g}^{r}(t,t^{\prime })$ is the retarded Green function for the
isolated quantum dots, and can be easily obtained as, 
\begin{equation}
{\bf g}^{r}(t,t^{\prime })=-i\theta (t-t^{\prime })\left( 
\begin{array}{cccc}
e^{-i\varepsilon _{1}(t-t^{\prime })} & 0 & 0 & 0 \\ 
0 & e^{i\varepsilon _{1}(t-t^{\prime })} & 0 & 0 \\ 
0 & 0 & e^{-i\varepsilon _{2}(t-t^{\prime })} & 0 \\ 
0 & 0 & 0 & e^{i\varepsilon _{2}(t-t^{\prime })}%
\end{array}%
\right) .  \eqnum{9}
\end{equation}%
Self-energy ${\bf \Sigma }^{r}(t_{1},t_{2})$ consists of three parts, such
that,

\begin{equation}
{\bf \Sigma }^{r}(t_{1},t_{2})={\bf \Sigma }_{12}+{\bf \Sigma }%
_{L}^{r}(t_{1},t_{2})+{\bf \Sigma }_{R}^{r}(t_{1},t_{2}),  \eqnum{10}
\end{equation}%
where ${\bf \Sigma }_{L}^{r}(t_{1},t_{2})$ results from the tunneling
coupling between QD and the normal metal lead, ${\bf \Sigma }%
_{R}^{r}(t_{1},t_{2})$ between QD and superconductor lead, and ${\bf \Sigma }%
_{12}$ is from interdot coupling contribution. Under the wide-bandwidth
approximation \cite{8,16,20}, the self-energy becomes 
\begin{equation}
{\bf \Sigma }_{12}=\left( 
\begin{array}{cccc}
0 & 0 & \Omega ^{\ast }e^{-i\Phi _{12}} & 0 \\ 
0 & 0 & 0 & -\Omega e^{i\Phi _{12}} \\ 
\Omega e^{i\Phi _{12}} & 0 & 0 & 0 \\ 
0 & -\Omega ^{\ast }e^{-i\Phi _{12}} & 0 & 0%
\end{array}%
\right) ,  \eqnum{11}
\end{equation}%
and ${\bf \Sigma }_{L}^{r}(t_{1},t_{2})$ is written as 
\begin{eqnarray}
{\bf \Sigma }_{L}^{r}(t_{1},t_{2}) &=&\sum_{k}{\bf L}^{\ast }{\bf g}%
_{k}^{r}(t_{1},t_{2}){\bf L}  \eqnum{12} \\
&=&-\frac{i}{2}\delta (t_{1}-t_{2}){\bf \Gamma }^{L},  \nonumber
\end{eqnarray}%
where 
\begin{equation}
{\bf L}=\left( 
\begin{array}{cccc}
L_{1}e^{-i\frac{\Phi }{4}} & 0 & 0 & 0 \\ 
0 & -L_{1}^{\ast }e^{i\frac{\Phi }{4}} & 0 & 0 \\ 
0 & 0 & L_{2}e^{i\frac{\Phi }{4}} & 0 \\ 
0 & 0 & 0 & -L_{2}^{\ast }e^{-i\frac{\Phi }{4}}%
\end{array}%
\right) .  \eqnum{13}
\end{equation}%
The self-energy from the coupling between QD and right superconductor lead
is given by \cite{16}, 
\begin{eqnarray}
&&\Sigma _{R}^{r}(t_{1},t_{2})  \nonumber \\
&=&\sum_{p}{\bf R}^{\ast }{\bf g}_{p}^{r}(t_{1},t_{2}){\bf R}  \nonumber \\
&=&-i\theta (t_{1}-t_{2})  \nonumber \\
&&\times \int \frac{d\varepsilon }{2\pi }\frac{e^{-i\varepsilon (t-t^{\prime
})}\left\vert \varepsilon \right\vert }{\sqrt{\varepsilon ^{2}-\Delta ^{2}}}%
\left( 
\begin{array}{cccc}
\Gamma _{1}^{R} & -\Gamma _{1}^{R}e^{-i\frac{\Phi }{2}}\frac{\Delta }{%
\left\vert \varepsilon \right\vert } & \sqrt{\Gamma _{1}^{R}\Gamma _{2}^{R}}%
e^{-i\frac{\Phi }{2}} & -\sqrt{\Gamma _{1}^{R}\Gamma _{2}^{R}}\frac{\Delta }{%
\left\vert \varepsilon \right\vert } \\ 
-\Gamma _{1}^{R}e^{i\frac{\Phi }{2}}\frac{\Delta }{\left\vert \varepsilon
\right\vert } & \Gamma _{1}^{R} & -\sqrt{\Gamma _{1}^{R}\Gamma _{2}^{R}}%
\frac{\Delta }{\left\vert \varepsilon \right\vert } & \sqrt{\Gamma
_{1}^{R}\Gamma _{2}^{R}}e^{i\frac{\Phi }{2}} \\ 
\sqrt{\Gamma _{1}^{R}\Gamma _{2}^{R}}e^{i\frac{\Phi }{2}} & -\sqrt{\Gamma
_{1}^{R}\Gamma _{2}^{R}}\frac{\Delta }{\left\vert \varepsilon \right\vert }
& \Gamma _{2}^{R} & -\Gamma _{2}^{R}e^{i\frac{\Phi }{2}}\frac{\Delta }{%
\left\vert \varepsilon \right\vert } \\ 
-\sqrt{\Gamma _{1}^{R}\Gamma _{2}^{R}}\frac{\Delta }{\left\vert \varepsilon
\right\vert } & \sqrt{\Gamma _{1}^{R}\Gamma _{2}^{R}}e^{-i\frac{\Phi }{2}} & 
-\Gamma _{2}^{R}e^{-i\frac{\Phi }{2}}\frac{\Delta }{\left\vert \varepsilon
\right\vert } & \Gamma _{2}^{R}%
\end{array}%
\right) ,  \nonumber \\
&&  \eqnum{14}
\end{eqnarray}%
where 
\begin{equation}
{\bf R}=\left( 
\begin{array}{cccc}
R_{1}e^{i\frac{\Phi }{4}} & 0 & 0 & 0 \\ 
0 & -R_{1}^{\ast }e^{-i\frac{\Phi }{4}} & 0 & 0 \\ 
0 & 0 & R_{2}e^{-i\frac{\Phi }{4}} & 0 \\ 
0 & 0 & 0 & -R_{2}^{\ast }e^{i\frac{\Phi }{4}}%
\end{array}%
\right) ,  \eqnum{15}
\end{equation}%
and the retarded Green function of the isolated superconductor lead $%
g_{p}^{r}(t,t^{\prime })$ is read as 
\begin{equation}
{\bf g}_{p}^{r}(t,t^{\prime })=-i\theta (t-t^{\prime })\int d\varepsilon
\rho _{R}^{N}\frac{e^{-i\varepsilon (t-t^{\prime })}\left\vert \varepsilon
\right\vert }{\sqrt{\varepsilon ^{2}-\Delta ^{2}}}\left( 
\begin{array}{cccc}
1 & \frac{\Delta }{\left\vert \varepsilon \right\vert } & 1 & \frac{\Delta }{%
\left\vert \varepsilon \right\vert } \\ 
\frac{\Delta }{\left\vert \varepsilon \right\vert } & 1 & \frac{\Delta }{%
\left\vert \varepsilon \right\vert } & 1 \\ 
1 & \frac{\Delta }{\left\vert \varepsilon \right\vert } & 1 & \frac{\Delta }{%
\left\vert \varepsilon \right\vert } \\ 
\frac{\Delta }{\left\vert \varepsilon \right\vert } & 1 & \frac{\Delta }{%
\left\vert \varepsilon \right\vert } & 1%
\end{array}%
\right) .  \eqnum{16}
\end{equation}%
with $\rho _{R}^{N}$ being the density states of the right lead in a normal
metal state. We define $\Gamma _{1}^{R}$ $=$ $2\pi \left\vert
R_{1}\right\vert ^{2}\rho _{R}^{N}$. By substituting Eqs. (9)--(13) into Eq.
(8), the Green function ${\bf G}^{r}(t,t^{\prime })$ can be derived. It is,
however, convenient to transform Eq. (8) into the energy representation by
making use of the Fourier transformation, 
\begin{equation}
{\bf G}^{r}(E,E^{\prime })=\int dtdt^{\prime }{\bf G}^{r}(t,t^{\prime
})e^{iEt}e^{-iE^{\prime }t^{\prime }}.  \eqnum{17}
\end{equation}%
Then solving Eq. (8) in the energy representation, we get (see Appendix A)

\begin{equation}
G_{ij}^{r}(E,E^{\prime })=2\pi \widetilde{G}_{ij}^{r}(E^{\prime })\delta
(E-E^{\prime }),  \eqnum{18}
\end{equation}%
here, $\ i,j=1,2,3,4$. $G_{ij}^{r}(E,E^{\prime })$ are the elements of the
matrix ${\bf G}^{r}(E,E^{\prime })$, and the detailed calculation of $%
\widetilde{G}_{ij}^{r}(E^{\prime })$ is given in Appendix A. The Green
function ${\bf G}^{r}(t,t^{\prime })$ is obtained after making an inverse
Fourier transformation as 
\begin{equation}
{\bf G}^{r}(t,t^{\prime })=\frac{1}{\left( 2\pi \right) ^{2}}\int
dEdE^{\prime }{\bf G}^{r}(E,E^{\prime })e^{-iEt}e^{iE^{\prime }t^{\prime }}.
\eqnum{19}
\end{equation}%
In the following, we derive the smaller Green function using the Keldysh
equation 
\begin{equation}
{\bf G}^{<}(t,t^{\prime })=\int dt_{1}dt_{2}{\bf G}^{r}(t,t_{1}){\bf \Sigma }%
^{<}(t_{1},t_{2}){\bf G}^{a}(t_{2},t^{\prime }),  \eqnum{20}
\end{equation}%
where, 
\begin{equation}
{\bf \Sigma }^{<}(t_{1},t_{2})={\bf \Sigma }_{L}^{<}(t_{1},t_{2})+{\bf %
\Sigma }_{R}^{<}(t_{1},t_{2}),  \eqnum{21}
\end{equation}%
with 
\begin{eqnarray}
&&{\bf \Sigma }_{L}^{<}(t_{1},t_{2})  \nonumber \\
&=&\sum_{k}{\bf L}^{\ast }{\bf g}_{k}^{<}(t_{1},t_{2}){\bf L}  \nonumber \\
&=&i\int \frac{d\varepsilon }{2\pi }e^{-i\varepsilon (t_{1}-t_{2})}\left( 
\begin{array}{cccc}
f_{L}(\varepsilon +eV) & 0 & 0 & 0 \\ 
0 & f_{L}(\varepsilon -eV) & 0 & 0 \\ 
0 & 0 & f_{L}(\varepsilon +eV) & 0 \\ 
0 & 0 & 0 & f_{L}(\varepsilon -eV)%
\end{array}%
\right) {\bf \Gamma }^{L},  \eqnum{22}
\end{eqnarray}%
and 
\begin{eqnarray}
{\bf \Sigma }_{R}^{<}(t_{1},t_{2}) &=&\sum_{p}{\bf R}^{\ast }{\bf g}%
_{p}^{<}(t_{1},t_{2}){\bf R}  \nonumber \\
&=&i\int \frac{d\varepsilon }{2\pi }e^{-i\varepsilon
(t_{1}-t_{2})}f_{R}(\varepsilon )\widetilde{\rho }_{R}(\varepsilon ) 
\nonumber \\
&&\times \left( 
\begin{array}{cccc}
\Gamma _{1}^{R} & -\Gamma _{1}^{R}e^{-i\frac{\Phi }{2}}\frac{\Delta }{%
\left\vert \varepsilon \right\vert } & \sqrt{\Gamma _{1}^{R}\Gamma _{2}^{R}}%
e^{-i\frac{\Phi }{2}} & -\sqrt{\Gamma _{1}^{R}\Gamma _{2}^{R}}\frac{\Delta }{%
\left\vert \varepsilon \right\vert } \\ 
-\Gamma _{1}^{R}e^{i\frac{\Phi }{2}}\frac{\Delta }{\left\vert \varepsilon
\right\vert } & \Gamma _{1}^{R} & -\sqrt{\Gamma _{1}^{R}\Gamma _{2}^{R}}%
\frac{\Delta }{\left\vert \varepsilon \right\vert } & \sqrt{\Gamma
_{1}^{R}\Gamma _{2}^{R}}e^{i\frac{\Phi }{2}} \\ 
\sqrt{\Gamma _{1}^{R}\Gamma _{2}^{R}}e^{i\frac{\Phi }{2}} & -\sqrt{\Gamma
_{1}^{R}\Gamma _{2}^{R}}\frac{\Delta }{\left\vert \varepsilon \right\vert }
& \Gamma _{2}^{R} & -\Gamma _{2}^{R}e^{i\frac{\Phi }{2}}\frac{\Delta }{%
\left\vert \varepsilon \right\vert } \\ 
-\sqrt{\Gamma _{1}^{R}\Gamma _{2}^{R}}\frac{\Delta }{\left\vert \varepsilon
\right\vert } & \sqrt{\Gamma _{1}^{R}\Gamma _{2}^{R}}e^{-i\frac{\Phi }{2}} & 
-\Gamma _{2}^{R}e^{-i\frac{\Phi }{2}}\frac{\Delta }{\left\vert \varepsilon
\right\vert } & \Gamma _{2}^{R}%
\end{array}%
\right) ,  \eqnum{23}
\end{eqnarray}%
where 
\begin{equation}
\widetilde{\rho }_{R}(\varepsilon )=\frac{\left\vert \varepsilon \right\vert
\theta (\left\vert \varepsilon \right\vert -\Delta )}{\sqrt{\varepsilon
^{2}-\Delta ^{2}}}.  \eqnum{24}
\end{equation}%
$\widetilde{\rho }_{R}(\varepsilon )$ is the ratio of superconducting
density of states to the normal density of states \cite{15}. According to
Eq. (20), ${\bf G}^{<}(t,t)$ can be obtained by substituting Eqs. (21)-(23)
and the Green functions ${\bf G}^{r}(t_{1},t_{2})${\bf \ , }${\bf G}%
^{a}(t_{1},t_{2})${\bf \ } into Eq. (20) with the results of matrix elements 
$G_{11}^{<}(t,t),G_{13}^{<}(t,t),G_{31}^{<}(t,t),$ and $G_{33}^{<}(t,t)$
shown in Appendix B.

The total current, i.e., the sum of both spin-up and spin-down parts, is
found by substituting the retarded Green function $G_{ij}^{r}(t,t^{\prime })$
and the smaller Green function $G_{ij}^{<}(t,t)$ shown in Appendix B into
Eq. (6), and the result is 
\begin{equation}
J_{L}=J_{A}+J_{T}  \nonumber
\end{equation}%
where 
\[
J_{A}=\frac{2e}{\hbar }\int \frac{d\varepsilon }{2\pi }[f_{L}(\varepsilon
+eV)-f_{L}(\varepsilon -eV)]T_{AR}, 
\]%
and 
\[
J_{T}=\frac{2e}{\hbar }\int \frac{d\varepsilon }{2\pi }[f_{L}(\varepsilon
+eV)-f_{R}(\varepsilon )]\widetilde{\rho }_{R}(\varepsilon )T_{LR}. 
\]%
$J_{A}$ is seen to be the current due to the Andreev reflection contribution
and the probability of the Andreev reflection $T_{AR}$ is given by 
\begin{eqnarray}
T_{AR} &=&(\Gamma _{1}^{L})^{2}\left\vert \widetilde{G}_{12}^{r}(\varepsilon
)\right\vert ^{2}+(\Gamma _{2}^{L})^{2}\left\vert \widetilde{G}%
_{34}^{r}(\varepsilon )\right\vert ^{2}  \nonumber \\
&&+\Gamma _{1}^{L}\Gamma _{2}^{L}(\left\vert \widetilde{G}%
_{14}^{r}(\varepsilon )\right\vert ^{2}+\left\vert \widetilde{G}%
_{32}^{r}(\varepsilon )\right\vert ^{2})  \nonumber \\
&&+2\sqrt{\Gamma _{1}^{L}\Gamma _{2}^{L}}%
\mathop{\rm Re}%
[\Gamma _{1}^{L}e^{-i\frac{\Phi }{2}}\widetilde{G}_{12}^{r}(\varepsilon )%
\widetilde{G}_{14}^{r\ast }(\varepsilon )  \nonumber \\
&&+\Gamma _{2}^{L}e^{-i\frac{\Phi }{2}}\widetilde{G}_{32}^{r}(\varepsilon )%
\widetilde{G}_{34}^{r\ast }(\varepsilon )+\Gamma _{1}^{L}e^{-i\frac{\Phi }{2}%
}\widetilde{G}_{12}^{r}(\varepsilon )\widetilde{G}_{32}^{r\ast }(\varepsilon
)  \eqnum{26} \\
&&+\sqrt{\Gamma _{1}^{L}\Gamma _{2}^{L}}e^{-i\Phi }\widetilde{G}%
_{12}^{r}(\varepsilon )\widetilde{G}_{34}^{r\ast }(\varepsilon )+\sqrt{%
\Gamma _{1}^{L}\Gamma _{2}^{L}}\widetilde{G}_{14}^{r}(\varepsilon )%
\widetilde{G}_{32}^{r\ast }(\varepsilon )  \nonumber \\
&&+\Gamma _{2}^{L}e^{-i\frac{\Phi }{2}}\widetilde{G}_{14}^{r}(\varepsilon )%
\widetilde{G}_{34}^{r\ast }(\varepsilon )].  \nonumber
\end{eqnarray}%
As explained in Refs. \cite{15,24}, the term $(\Gamma
_{1}^{L})^{2}\left\vert \widetilde{G}_{12}^{r}(\varepsilon )\right\vert ^{2}$
[$(\Gamma _{2}^{L})^{2}\left\vert \widetilde{G}_{34}^{r}(\varepsilon
)\right\vert ^{2}$]{\bf \ d}escribes an electron with spin-up tunnelling
from the left lead through quantum dot 1 (2) to the superconductor lead and
reflecting a hole back to the left lead through the quantum dot 1 (2). While
the term{\bf \ }$\Gamma _{1}^{L}\Gamma _{2}^{L}\left\vert \widetilde{G}%
_{14}^{r}(\varepsilon )\right\vert ^{2}${\bf [}$\Gamma _{1}^{L}\Gamma
_{2}^{L}\left\vert \widetilde{G}_{32}^{r}(\varepsilon )\right\vert ^{2}${\bf %
] }is for the spin-up electron tunneling from left lead through dot 1 (2) to
the superconductor lead and a hole reflecting back to the left lead through
the dot 2 (1). There exist other channels in which the tunnel electron and
reflecting hole go through both two dots. \ The remaining part in Eq. (26)
describes the interference effect of all possible multichannels of the
device. The transmission probability from left to right leads reads

\begin{eqnarray}
T_{LR} &=&\Gamma _{1}^{L}[\Gamma _{1}^{R}\left\vert \widetilde{G}%
_{11}^{r}(\varepsilon )\right\vert ^{2}+\Gamma _{1}^{R}\left\vert \widetilde{%
G}_{12}^{r}(\varepsilon )\right\vert ^{2}  \nonumber \\
&&+\Gamma _{2}^{R}\left\vert \widetilde{G}_{13}^{r}(\varepsilon )\right\vert
^{2}+\Gamma _{2}^{R}\left\vert \widetilde{G}_{14}^{r}(\varepsilon
)\right\vert ^{2}]  \nonumber \\
&&+\Gamma _{2}^{L}[\Gamma _{1}^{R}\left\vert \widetilde{G}%
_{31}^{r}(\varepsilon )\right\vert ^{2}+\Gamma _{1}^{R}\left\vert \widetilde{%
G}_{32}^{r}(\varepsilon )\right\vert ^{2}  \eqnum{27} \\
&&+\Gamma _{2}^{R}\left\vert \widetilde{G}_{33}^{r}(\varepsilon )\right\vert
^{2}+\Gamma _{2}^{R}\left\vert \widetilde{G}_{34}^{r}(\varepsilon
)\right\vert ^{2}]  \nonumber \\
&&+2%
\mathop{\rm Re}%
[-\Gamma _{1}^{L}\Gamma _{1}^{R}e^{-i\frac{\Phi }{2}}\frac{\Delta }{%
\left\vert \varepsilon \right\vert }\widetilde{G}_{11}^{r}(\varepsilon )%
\widetilde{G}_{12}^{r\ast }(\varepsilon )+\cdots ].  \nonumber
\end{eqnarray}%
containing many terms which are understood. We are mainly interested in the
physics of the Andreev reflections. As a matter of fact only the Andreev
reflection process contributes to the current (i. e., $J_{T}=0$) at zero
temperature, if $\left\vert eV\right\vert <\Delta .$ In the following
section, we analyze time evolution of the probability of the Andreev
reflection $T_{AR}$ and the Andreev reflection current $J_{A}$ as well
depending on dc bias voltage, gate voltage, magnetic flux, and interdot
coupling strength.

\section{NUMERICAL RESULTS}

First of all we transform the coupled-QD system into an effective decoupled
one with bonding and antibonding dressed states of the QD molecule using the
following transformation \cite{23},

\begin{equation}
\left( 
\begin{array}{c}
f_{+} \\ 
f_{-}%
\end{array}%
\right) =\left( 
\begin{array}{cc}
-\cos \beta e^{i\Phi _{12}} & -\sin \beta \\ 
-\sin \beta & \cos \beta e^{-i\Phi _{12}}%
\end{array}%
\right) \left( 
\begin{array}{c}
d_{1} \\ 
d_{2}%
\end{array}%
\right) .  \eqnum{28}
\end{equation}%
where, $f_{-}$ and $f_{+}$ are the annihilation operators of the bonding and
antibonding dressed states for the QD molecule. And $\beta $ is defined as
1/2tan$^{-1}$[2$\Omega $/($\varepsilon _{1}$-$\varepsilon _{2}$)]. Thus the
Hamiltonian for coupled double quantum dot $H_{D}$ is diagonalized as 
\begin{equation}
\widetilde{H}_{D}=E_{+}f_{+}^{\dagger }f_{+}+E_{-}f_{-}^{\dagger }f_{-}, 
\eqnum{29}
\end{equation}%
where 
\begin{equation}
E_{\pm }=\frac{1}{2}[\varepsilon _{1}+\varepsilon _{2}\pm \sqrt{(\varepsilon
_{1}-\varepsilon _{2})^{2}+4\Omega ^{2}}].  \eqnum{30}
\end{equation}%
Considering the tunneling Hamiltonian between double QD and the left lead,
we obtain the linewidth $\Gamma _{+}^{L}$ and $\Gamma _{-}^{L}$
corresponding to the antibonding and bonding states coupled to the left
lead, respectively, as

\begin{eqnarray}
\Gamma _{+}^L &=&\Gamma _1^L\cos ^2\beta +\Gamma _2^L\sin ^2\beta +\sqrt{%
\Gamma _1^L\Gamma _2^L}\sin (2\beta )\cos (\frac \Phi 2+\Phi _{12}), 
\eqnum{31} \\
\Gamma _{-}^L &=&\Gamma _1^L\sin ^2\beta +\Gamma _2^L\cos ^2\beta -\sqrt{%
\Gamma _1^L\Gamma _2^L}\sin (2\beta )\cos (\frac \Phi 2+\Phi _{12}). 
\eqnum{32}
\end{eqnarray}

In the numerical analysis we set $\Delta =1$ as the energy unit. The energy
dependence of reflection probability $T_{AR}$ on the interdot coupling
strength is shown in Fig. $2$ for the symmetry [with $\Gamma _{1}^{L}=\Gamma
_{1}^{R}=$ $\Gamma _{2}^{L}=\Gamma _{2}^{R}=0.02$ ,Fig.2(a)] and asymmetry
[with $\Gamma _{1}^{L}=\Gamma _{2}^{L}=0.08$, $\Gamma _{1}^{R}=\Gamma
_{2}^{R}=0.02$ ,Fig.2(b) and $\Gamma _{1}^{L}=$ $\Gamma _{1}^{R}=0.08$, $%
\Gamma _{2}^{L}=\Gamma _{2}^{R}=0.02$ ,Fig.2(c)] cases, respectively. Other
parameters are chosen as $\varepsilon _{1}=\varepsilon _{2}=0$, and $\Phi
=\Phi _{12}=0.$ It is seen in Fig. $2(a)$ that if the two QDs are completely
decoupled (i.e., $\Omega =0$) only a single peak appears (solid line), and
when $\Omega \neq 0$ , two peaks emerges at $\varepsilon =\pm \Omega $ due
to the existence of two energy levels. The maximum of peaks is getting lower
when the coupling constant $\Omega $ is increasing.{\bf \ }For the asymmetry
case Fig. $2(b)$\ shows that there are no significant two peaks. Fig. $2(c)$%
\ exhibits the complex spectra due to the possible overlap of the broad
levels similar to the interpretation of Fano effect in a parallel coupled
double QD connected to two normal metal leads system \cite{23}. And when $%
\Omega $\ is large enough such that the broad two energy levels are
completely separated, the two single peaks remain at $\varepsilon =E_{+}$\
and $E_{-}$.

It is interesting to see the effect of the magnetic flux on the Andreev
reflection and Fig. $3$ presents the dependence of the Andreev reflection
probability on the magnetic flux $\Phi $ with $\alpha _{1}=\alpha _{2}$ and
therefore $\Phi _{12}=0$. The height of peaks at $\varepsilon =\pm \Omega $
can reach $\ 2$ when the magnetic flux is nonzero, and magnetic flux also
has an influence on the width of the peaks, i. e., on the linewidth function 
$\Gamma _{\pm }$, here we choose $\Phi =\pi /3$\ for a solid line, and $2\pi
/3$\ for a dotted line, the height of peaks are not changed. Electron phase
coherence is shown in Fig. $4$. When $\Omega =0$, i. e., there is no the
interdot coupling, the oscillation period of the Andreev reflection
probability vs magnetic flux is $2\pi \Phi _{0}$ (solid line), while $\Omega
\neq 0$, the period changes to $4\pi \Phi _{0}$when $\alpha _{1}=\alpha _{2}$
[see Fig. $4(a)$]. Here, we choose the different energy eigenvalues such
that $\varepsilon =0.11$, since at those values the probability of the
Andreev reflection reaches maximum. To see the reason of the period
changing, we give a general result of the oscillation period when $\alpha
_{1}/\alpha _{2}=n$. The channel paths contain two parts due to the Andreev
reflection in the hybrid system. One is the incident electron from the left
lead to the superconductor in Fig . $5a$, the other is the reflecting hole
from the superconductor to the left lead in Fig. $5b$. The phase shift of
electron in the case (3) in Fig. $5a$\ is $(\Phi _{1}-\Phi _{2})/2$, and the
corresponding phase shift of hole in the case (1) in Fig. $5b$ is $(\Phi
_{1}+\Phi _{2})/2$, so the total shift of the channel (3) for the elcetron
and (1) for the hole is $\Phi _{1}$. The interference effect corresponds to
all the channel paths reads \cite{25} 
\begin{eqnarray}
T &=&\left\vert 
\begin{array}{c}
A_{0}+A_{1}e^{i\Phi _{1}}+A_{2}e^{-i\Phi _{1}}+A_{3}e^{i\Phi
_{2}}+A_{4}e^{-i\Phi _{2}}+A_{5}e^{i(\Phi _{1}+\Phi _{2})} \\ 
+A_{6}e^{-i(\Phi _{1}+\Phi _{2})}+A_{7}e^{i(\Phi _{1}-\Phi
_{2})}+A_{8}e^{-i(\Phi _{1}-\Phi _{2})}%
\end{array}%
\right\vert ^{2}  \nonumber \\
&=&B_{0}+B_{1}\cos \Phi _{1}+B_{2}\cos \Phi _{2}+B_{3}\cos (\Phi _{1}+\Phi
_{2})  \nonumber \\
&&+B_{4}\cos (\Phi _{1}-\Phi _{2})+B_{5}\cos 2\Phi _{1}+B_{6}\cos 2\Phi _{2}
\nonumber \\
&&+B_{7}\cos 2(\Phi _{1}+\Phi _{2})+B_{8}\cos (2\Phi _{1}+\Phi
_{2})+B_{9}\cos (\Phi _{1}+2\Phi _{2})  \nonumber \\
&&+B_{10}\cos 2(\Phi _{1}-\Phi _{2})+B_{11}\cos (2\Phi _{1}-\Phi
_{2})+B_{12}\cos (\Phi _{1}-2\Phi _{2})  \nonumber \\
&&  \eqnum{33}
\end{eqnarray}%
where parameters $A$\ are the amplitudes of channel paths. With $\Phi
_{1}=\Phi n/(n+1)$,$\Phi _{2}$ $=\Phi /(n+1)$, Eq. (33) shows the
oscillation periods might have $\pi $, $2\pi $, $2\pi (n+1)/n$, $2\pi (n+1)$%
, $2\pi (n+1)/2n$, $2\pi (n+1)/2$, $2\pi (n+1)/(n-1)$, $2\pi (n+1)/(n-2)$, $%
2\pi (n+1)/(n+2)$, $2\pi (n+1)/(2n+1)$, and $2\pi (n+1)/(2n-1)$, suggesting
that the oscillation period of reflect probability should be $2(n+1)\pi $.
The linear depedence of the ratio $n$\ is further confirmed in Fig. 4(b), it
shows that the oscillation period is $6\pi $ for $n=2$\ (solid line), $8\pi $%
\ for $n=3$\ (dotted line), and $10\pi $\ for $n=4$\ (dashed line).\ The
linear manner of the osicllation period is the same as the one in Ref. 22,
however, the mechanism is different: there the transmitted current is
analyzed.

Fig. $6$ shows current-gate-voltage curve ($J_{A}$ vs $V_{g}$) for the case
that $\varepsilon _{1}=0.1$, $\varepsilon _{2}=0.3$, $\Phi =\Phi _{12}=0$,
and the bias voltage $V=0.4$. We assume that the energy levels in both dots
are connected to the same voltage through the capacitive coupling. When $%
\Omega =0$ (solid line), there exist two smaller peaks at $V_{g}=0.1$, and $%
0.3$ located symmetrically beside the larger peak at $V_{g}=0.2$ with the
height that is two times higher than the height of the smaller peaks.{\bf \ }%
The interpretation of the phenomena is similar to that in Ref. \cite{15}
such that when $V_{g}=0.1$\ the energy level $\varepsilon _{1}$\ matches
with chemical potential $u_{R}=0$\ in the right superconductor lead and the
Andreev reflection takes place at this level. When $V_{g}=0.2$\ two energy
levels have a symmetric localization about $u_{R}$\ ( $\varepsilon _{2}$\ is 
$0.1$\ above $u_{R}$\ while $\varepsilon _{1}$\ is $0.1$\ below $u_{R}$\ )
and both $\varepsilon _{1}$\ and $\varepsilon _{2}$\ are below the bias
voltage, so the Andreev reflection that involves those two levels leads to
the height of the peaks that are two times higher than that containing only
one energy level{\bf .} The interpretation is the same for the gate voltage $%
V_{g}=0.3$ where the energy level $\varepsilon _{2}$ matches with $u_{R}$.
Some new properties are observed for the coupled case $\Omega \neq 0$: (1)
the location of the middle peak does not change, while the positions of the
left and right peaks shift towards the left and right sides, respectively,
with the increasing coupling strength $\Omega $. (2) the heights of the left
and the middle peaks are suppressed significantly; meantime, the height of
the right peak is enhanced. (3) the widths of the left and middle peaks
decrease while the width of the right peak widened. These interesting
phenomena can be explained with Eq. (30)-(32). The energy level of dressed
state $E_{+}$ gets larger and $E_{-}$ becomes smaller with increasing
coupling strength $\Omega $ , and at the end $E_{+}$ is higher than energy
level of bared state $\varepsilon _{2}$, when $E_{-}$ is smaller than $%
\varepsilon _{1}$. So when the gate voltage takes action on the dressed
states, the left peak runs towards the left while the right peak goes
opposite, however, the locations of the two peaks remain symmetric about the
middle peak. We assume $E_{-}=$ $\varepsilon _{1}-a=0.1-a$, $%
E_{+}=\varepsilon _{2}+a=0.3+a$, where $a$ is introduced due to the coupling
strength $\Omega $. So when the gate voltage becomes $V_{g}=0.1-a$, the left
peak emerges and with increasing $V_{g}$ (up to $0.2$), both two dressed
states levels locate symmetrically about $u_{R}$. Thus, the location of
middle peak does not change at $V_{g}=0.2$, no matter how the parameter $%
a(\Omega )$ varies. The linewidth of dressed states, $\Gamma _{+}^{L}$ ,
increases with the increasing of $\Omega $, while $\Gamma _{-}^{L}$
decreases under the parameters considered here, and hence, the right peak
gets wider and left peak becomes narrower with the increasing of $\Omega $.
The linewidths $\Gamma _{\pm }$ are directly connected with the coupling
strengths between QD and the leads. The widening of the right-side peak
means the increasing of coupling strength between dressed state $E_{+}$ and
the leads. Therefore, the height of the right-side peak is enhanced, and
opposite to this, the height of the left-side peak is suppressed. The
suppression of the middle-peak height suggests that the Andreev resonance
involving two energy levels gets maximum when two energy levels are
symmetric.

Figures. $7-9$ show a tunable Andreev reflection current depending on the
magnetic flux when $\Omega \neq 0$. In Fig. $7$, we fix $\Phi =\pi /2$, and
tune the value of $\Phi _{12}$ for different curves. One can see that the
height of the right peak is enhanced and the height of the left one is
suppressed while the height of the middle peak does not change with the
increase of $\Phi _{12}$. It is interesting to notice that the left peaks
has essentially no changes when $\Phi _{12}=0$ (solid line) and $\pi /2$
(dotted line), while the right peak changes conspicuously with respect to
the case without the magnetic flux. Figure $8$ has the opposite effect
compared with Fig. $7$, namely, the height of the left peak is suppressed
while the height of the right one is enhanced and the middle peak does not
vary with $\Phi _{12}$. We fix $\Phi _{12}=\pi /5$, and vary values of
magnetic flux $\Phi $ in Fig. $9$, it is shown that the height of three
peaks can be tuned simultaneously.

\section{CONCLUSIONS}

We have studied the tunable Andreev reflection in a double AB interferometer
in terms of the nonequilibrium Green function method and observed several
features of the reflection current. It is found that the oscillation period
of the Andreev reflection probability with the magnetic flux is $2\pi $ when
interdot coupling vanishes ( $\Omega =0$, in this case the system reduces to
a single interferometer), while it is $2(n+1)\pi $\ for our double
interferometer ( $\Omega \neq 0$) where $n$\ is the ratio of two parts
magnetic fluxes, i.e. $n=\Phi _{1}/\Phi _{2}$.{\bf \ }The Andreev reflection
current peaks cannot only be tuned by coupling strength $\Omega $, but also
by magnetic flux.

\section{ACKNOWLEDGMENT}

This work was supported by the National Natural Science Foundation of China
under Grant No. 10475053

\section{APPENDIX A}

In this appendix, we present the derivation of the Green function of Eq.
(18) in detail. From Eqs. (8)-(13) and (17) we have the following formulas
for the matrix elements of Green function $G^{r}(t,t^{\prime })$ in the
energy representation:

\begin{eqnarray}
g_{11}^{r-1}(E^{\prime })G_{11}^r(E,E^{\prime }) &=&2\pi \delta (E-E^{\prime
})-A_1G_{11}^r(E,E^{\prime })  \nonumber \\
&&+B\Gamma _1^Re^{i\frac \Phi 2}G_{12}^r(E,E^{\prime
})+x_1G_{13}^r(E,E^{\prime })  \eqnum{A1} \\
&&+B\sqrt{\Gamma _1^R\Gamma _2^R}G_{14}^r(E,E^{\prime }),  \nonumber
\end{eqnarray}
\begin{eqnarray}
g_{22}^{r-1}(E^{\prime })G_{12}^r(E,E^{\prime }) &=&B\Gamma _1^Re^{-i\frac %
\Phi 2}G_{11}^r(E,E^{\prime })-A_1G_{12}^r(E,E^{\prime })  \nonumber \\
&&+B\sqrt{\Gamma _1^R\Gamma _2^R}G_{13}^r(E,E^{\prime
})+x_2G_{14}^r(E,E^{\prime }),  \eqnum{A2}
\end{eqnarray}
\begin{eqnarray}
g_{33}^{r-1}(E^{\prime })G_{13}^r(E,E^{\prime }) &=&x_3G_{11}^r(E,E^{\prime
})+B\sqrt{\Gamma _1^R\Gamma _2^R}G_{12}^r(E,E^{\prime })  \nonumber \\
&&-A_2G_{13}^r(E,E^{\prime })+B\Gamma _2^Re^{-i\frac \Phi 2%
}G_{14}^r(E,E^{\prime }),  \eqnum{A3}
\end{eqnarray}
\begin{eqnarray}
g_{44}^{r-1}(E^{\prime })G_{14}^r(E,E^{\prime }) &=&B\sqrt{\Gamma _1^R\Gamma
_2^R}G_{11}^r(E,E^{\prime })+x_1G_{12}^r(E,E^{\prime })  \nonumber \\
&&+B\Gamma _2^Re^{i\frac \Phi 2}G_{13}^r(E,E^{\prime
})-A_2G_{14}^r(E,E^{\prime }),  \eqnum{A4}
\end{eqnarray}
where 
\begin{eqnarray}
A_1 &=&\frac i2(\Gamma _1^L+\Gamma _1^R\frac{\left| E^{\prime }\right| }{%
\sqrt{(E^{\prime })^2-\Delta ^2}}),  \nonumber \\
A_2 &=&\frac i2(\Gamma _2^L+\Gamma _2^R\frac{\left| E^{\prime }\right| }{%
\sqrt{(E^{^{\prime }})^2-\Delta ^2}}),  \eqnum{A5} \\
B &=&\frac i2\frac \Delta {\sqrt{(E^{^{\prime }})^2-\Delta ^2}},  \nonumber
\end{eqnarray}
and 
\begin{eqnarray}
x_1 &=&\Omega e^{i\Phi _{12}}-\frac i2\sqrt{\Gamma _1^L\Gamma _2^L}e^{-i%
\frac \Phi 2}-\frac i2\frac{\left| E^{\prime }\right| }{\sqrt{(E^{\prime
})^2-\Delta ^2}}\sqrt{\Gamma _1^R\Gamma _2^R}e^{i\frac \Phi 2},  \nonumber \\
x_2 &=&-\Omega ^{*}e^{-i\Phi _{12}}-\frac i2\sqrt{\Gamma _1^L\Gamma _2^L}e^{i%
\frac \Phi 2}-\frac i2\frac{\left| E^{\prime }\right| }{\sqrt{(E^{\prime
})^2-\Delta ^2}}\sqrt{\Gamma _1^R\Gamma _2^R}e^{-i\frac \Phi 2},  \nonumber
\\
x_3 &=&\Omega ^{*}e^{-i\Phi _{12}}-\frac i2\sqrt{\Gamma _1^L\Gamma _2^L}e^{i%
\frac \Phi 2}-\frac i2\frac{\left| E^{\prime }\right| }{\sqrt{(E^{^{\prime
}})^2-\Delta ^2}}\sqrt{\Gamma _1^R\Gamma _2^R}e^{-i\frac \Phi 2},  \eqnum{A6}
\\
x_4 &=&-\Omega e^{i\Phi _{12}}-\frac i2\sqrt{\Gamma _1^L\Gamma _2^L}e^{-i%
\frac \Phi 2}-\frac i2\frac{\left| E^{\prime }\right| }{\sqrt{(E^{^{\prime
}})^2-\Delta ^2}}\sqrt{\Gamma _1^R\Gamma _2^R}e^{i\frac \Phi 2}.  \nonumber
\end{eqnarray}

Solving those coupled equations, the results are obtained as 
\begin{eqnarray}
G_{11}^r(E,E^{\prime }) &=&2\pi \widetilde{G}_{11}^r(E^{\prime })\delta
(E-E^{\prime }),  \nonumber \\
G_{12}^r(E,E^{\prime }) &=&2\pi \widetilde{G}_{12}^r(E^{\prime })\delta
(E-E^{\prime }),  \nonumber \\
G_{13}^r(E,E^{\prime }) &=&2\pi \widetilde{G}_{13}^r(E^{\prime })\delta
(E-E^{\prime }),  \eqnum{A7} \\
G_{14}^r(E,E^{\prime }) &=&2\pi \widetilde{G}_{14}^r(E^{\prime })\delta
(E-E^{\prime }),  \nonumber
\end{eqnarray}
where 
\begin{equation}
\widetilde{G}_{11}^r(E^{\prime })=\frac 1{M(E^{\prime })}\left| 
\begin{array}{ccc}
g_{22}^{r-1}(E^{\prime })+A_1 & -B\sqrt{\Gamma _1^R\Gamma _2^R} & -x_2 \\ 
-B\sqrt{\Gamma _1^R\Gamma _2^R} & g_{33}^{r-1}(E^{\prime })+A_2 & -B\Gamma
_2^Re^{-i\frac \Phi 2} \\ 
-x_4 & -B\Gamma _2^Re^{i\frac \Phi 2} & g_{44}^{r-1}(E^{\prime })+A_2%
\end{array}
\right| ,  \eqnum{A8}
\end{equation}

\begin{equation}
\widetilde{G}_{12}^r(E^{\prime })=-\frac 1{M(E^{\prime })}\left| 
\begin{array}{ccc}
-B\Gamma _1^Re^{-i\frac \Phi 2} & -B\sqrt{\Gamma _1^R\Gamma _2^R} & -x_2 \\ 
-x_3 & g_{33}^{r-1}(E^{\prime })+A_2 & -B\Gamma _2^Re^{-i\frac \Phi 2} \\ 
-B\sqrt{\Gamma _1^R\Gamma _2^R} & -B\Gamma _2^Re^{i\frac \Phi 2} & 
g_{44}^{r-1}(E^{\prime })+A_2%
\end{array}
\right| ,  \eqnum{A9}
\end{equation}

\begin{equation}
\widetilde{G}_{13}^r(E^{\prime })=\frac 1{M(E^{\prime })}\left| 
\begin{array}{ccc}
-B\Gamma _1^Re^{-i\frac \Phi 2} & g_{22}^{r-1}(E^{\prime })+A_1 & -x_2 \\ 
-x_3 & -B\sqrt{\Gamma _1^R\Gamma _2^R} & -B\Gamma _2^Re^{-i\frac \Phi 2} \\ 
-B\sqrt{\Gamma _1^R\Gamma _2^R} & -x_4 & g_{44}^{r-1}(E^{\prime })+A_2%
\end{array}
\right| ,  \eqnum{A10}
\end{equation}

\begin{equation}
\widetilde{G}_{14}^r(E^{\prime })=-\frac 1{M(E^{\prime })}\left| 
\begin{array}{ccc}
-B\Gamma _1^Re^{-i\frac \Phi 2} & g_{22}^{r-1}(E^{\prime })+A_1 & -B\sqrt{%
\Gamma _1^R\Gamma _2^R} \\ 
-x_3 & -B\sqrt{\Gamma _1^R\Gamma _2^R} & g_{33}^{r-1}(E^{\prime })+A_2 \\ 
-B\sqrt{\Gamma _1^R\Gamma _2^R} & -x_4 & -B\Gamma _2^Re^{i\frac \Phi 2}%
\end{array}
\right| ,  \eqnum{A11}
\end{equation}
here, 
\begin{equation}
M(E^{\prime })=\left| 
\begin{array}{cccc}
g_{11}^{r-1}(E^{\prime })+A_1 & -B\Gamma _1^Re^{i\frac \Phi 2} & -x_1 & -B%
\sqrt{\Gamma _1^R\Gamma _2^R} \\ 
-B\Gamma _1^Re^{-i\frac \Phi 2} & g_{22}^{r-1}(E^{\prime })+A_1 & -B\sqrt{%
\Gamma _1^R\Gamma _2^R} & -x_2 \\ 
-x_3 & -B\sqrt{\Gamma _1^R\Gamma _2^R} & g_{33}^{r-1}(E^{\prime })+A_2 & 
-B\Gamma _2^Re^{-i\frac \Phi 2} \\ 
-B\sqrt{\Gamma _1^R\Gamma _2^R} & -x_4 & -B\Gamma _2^Re^{i\frac \Phi 2} & 
g_{44}^{r-1}(E^{\prime })+A_2%
\end{array}
\right| .  \eqnum{A12}
\end{equation}
With the same procedure we have

\begin{eqnarray}
g_{11}^{r-1}(E^{\prime })G_{31}^r(E,E^{\prime }) &=&-A_1G_{31}^r(E,E^{\prime
})+B\Gamma _1^Re^{i\frac \Phi 2}G_{32}^r(E,E^{\prime })  \nonumber \\
&&x_1G_{33}^r(E,E^{\prime })+B\sqrt{\Gamma _1^R\Gamma _2^R}%
G_{34}^r(E,E^{\prime }),  \eqnum{A13}
\end{eqnarray}

\begin{eqnarray}
g_{22}^{r-1}(E^{\prime })G_{32}^r(E,E^{\prime }) &=&B\Gamma _1^Re^{-i\frac %
\Phi 2}G_{31}^r(E,E^{\prime })-A_1G_{32}^r(E,E^{\prime })  \nonumber \\
&&+B\sqrt{\Gamma _1^R\Gamma _2^R}G_{33}^r(E,E^{\prime
})+x_2G_{34}^r(E,E^{\prime }),  \eqnum{A14}
\end{eqnarray}

\begin{eqnarray}
g_{33}^{r-1}(E^{\prime })G_{33}^r(E,E^{\prime }) &=&2\pi \delta (E-E^{\prime
})+x_3G_{31}^r(E,E^{\prime })  \nonumber \\
&&+B\sqrt{\Gamma _1^R\Gamma _2^R}G_{32}^r(E,E^{\prime
})-A_2G_{33}^r(E,E^{\prime })  \eqnum{A15} \\
&&+B\Gamma _2^Re^{-i\frac \Phi 2}G_{34}^r(E,E^{\prime }),  \nonumber
\end{eqnarray}

\begin{eqnarray}
g_{44}^{r-1}(E^{\prime })G_{34}^r(E,E^{\prime }) &=&B\sqrt{\Gamma _1^R\Gamma
_2^R}G_{31}^r(E,E^{\prime })+x_4G_{32}^r(E,E^{\prime })  \nonumber \\
&&+B\Gamma _2^Re^{i\frac \Phi 2}G_{33}^r(E,E^{\prime
})-A_2G_{44}^r(E,E^{\prime }).  \eqnum{A16}
\end{eqnarray}
So the final results are 
\begin{eqnarray}
G_{31}^r(E,E^{\prime }) &=&2\pi \widetilde{G}_{31}^r(E^{\prime })\delta
(E-E^{\prime })  \nonumber \\
G_{32}^r(E,E^{\prime }) &=&2\pi \widetilde{G}_{32}^r(E^{\prime })\delta
(E-E^{\prime })  \nonumber \\
G_{33}^r(E,E^{\prime }) &=&2\pi \widetilde{G}_{33}^r(E^{\prime })\delta
(E-E^{\prime })  \eqnum{A17} \\
G_{34}^r(E,E^{\prime }) &=&2\pi \widetilde{G}_{34}^r(E^{\prime })\delta
(E-E^{\prime })  \nonumber
\end{eqnarray}
where 
\begin{equation}
\widetilde{G}_{31}^r(E^{\prime })=\frac 1{M(E^{\prime })}\left| 
\begin{array}{ccc}
-B\Gamma _1^Re^{i\frac \Phi 2} & -x_1 & -B\sqrt{\Gamma _1^R\Gamma _2^R} \\ 
g_{22}^{r-1}(E^{\prime })+A_1 & -B\sqrt{\Gamma _1^R\Gamma _2^R} & -x_2 \\ 
-x_4 & -B\Gamma _2^Re^{i\frac \Phi 2} & g_{44}^{r-1}(E^{\prime })+A_2%
\end{array}
\right|  \eqnum{A18}
\end{equation}

\begin{equation}
\widetilde{G}_{32}^r(E^{\prime })=-\frac 1{M(E^{\prime })}\left| 
\begin{array}{ccc}
g_{11}^{r-1}(E^{\prime })+A_1 & -x_1 & -B\sqrt{\Gamma _1^R\Gamma _2^R} \\ 
-B\Gamma _1^Re^{-i\frac \Phi 2} & -B\sqrt{\Gamma _1^R\Gamma _2^R} & -x_2 \\ 
-B\sqrt{\Gamma _1^R\Gamma _2^R} & -B\Gamma _2^Re^{i\frac \Phi 2} & 
g_{44}^{r-1}(E^{\prime })+A_2%
\end{array}
\right|  \eqnum{A19}
\end{equation}

\begin{equation}
\widetilde{G}_{33}^r(E^{\prime })=\frac 1{M(E^{\prime })}\left| 
\begin{array}{ccc}
g_{11}^{r-1}(E^{\prime })+A_1 & -B\Gamma _1^Re^{i\frac \Phi 2} & -B\sqrt{%
\Gamma _1^R\Gamma _2^R} \\ 
-B\Gamma _1^Re^{-i\frac \Phi 2} & g_{22}^{r-1}(E^{\prime })+A_1 & -x_2 \\ 
-B\sqrt{\Gamma _1^R\Gamma _2^R} & -x_4 & g_{44}^{r-1}(E^{\prime })+A_2%
\end{array}
\right|  \eqnum{A20}
\end{equation}

\begin{equation}
\widetilde{G}_{34}^r(E^{\prime })=-\frac 1{M(E^{\prime })}\left| 
\begin{array}{ccc}
g_{11}^{r-1}(E^{\prime })+A_1 & -B\Gamma _1^Re^{i\frac \Phi 2} & -x_1 \\ 
-B\Gamma _1^Re^{-i\frac \Phi 2} & g_{22}^{r-1}(E^{\prime })+A_1 & -B\sqrt{%
\Gamma _1^R\Gamma _2^R} \\ 
-B\sqrt{\Gamma _1^R\Gamma _2^R} & -x_4 & -B\Gamma _2^Re^{i\frac \Phi 2}%
\end{array}
\right|  \eqnum{A21}
\end{equation}

\section{APPENDIX B}

The complete expression of matrix elements of the Green function $G^{<}(t,t) 
$ is 
\begin{eqnarray}
G_{11}^{<}(t,t) &=&i\int \frac{d\varepsilon }{2\pi }\{f_L(\varepsilon
+eV)\Gamma _1^L\left| \widetilde{G}_{11}^r(\varepsilon )\right|
^2+f_L(\varepsilon -eV)\Gamma _1^L\left| \widetilde{G}_{12}^r(\varepsilon
)\right| ^2  \nonumber \\
&&+f_L(\varepsilon +eV)\Gamma _2^L\left| \widetilde{G}_{13}^r(\varepsilon
)\right| ^2+f_L(\varepsilon -eV)\Gamma _2^L\left| \widetilde{G}%
_{14}^r(\varepsilon )\right| ^2  \nonumber \\
&&+2\sqrt{\Gamma _1^L\Gamma _2^L}%
\mathop{\rm Re}
[f_L(\varepsilon +eV)e^{i\frac \Phi 2}\widetilde{G}_{11}^r(\varepsilon )%
\widetilde{G}_{13}^{r*}(\varepsilon )  \nonumber \\
&&+f_L(\varepsilon -eV)e^{-i\frac \Phi 2}\widetilde{G}_{12}^r(\varepsilon )%
\widetilde{G}_{14}^{r*}(\varepsilon )]\}  \nonumber \\
&&+i\int \frac{d\varepsilon }{2\pi }f_R(\varepsilon )\widetilde{\rho }%
_R(\varepsilon )\{\Gamma _1^R\left| \widetilde{G}_{11}^r(\varepsilon
)\right| ^2+\Gamma _1^R\left| \widetilde{G}_{12}^r(\varepsilon )\right| ^2 
\nonumber \\
&&+\Gamma _2^R\left| \widetilde{G}_{13}^r(\varepsilon )\right| ^2+\Gamma
_2^R\left| \widetilde{G}_{14}^r(\varepsilon )\right| ^2  \eqnum{B1} \\
&&+2%
\mathop{\rm Re}
[-\Gamma _1^Re^{-i\frac \Phi 2}\frac \Delta {\left| \varepsilon \right| }%
\widetilde{G}_{11}^r(\varepsilon )\widetilde{G}_{12}^{r*}(\varepsilon ) 
\nonumber \\
&&+\sqrt{\Gamma _1^R\Gamma _2^R}e^{-i\frac \Phi 2}\widetilde{G}%
_{11}^r(\varepsilon )\widetilde{G}_{13}^{r*}(\varepsilon )+\sqrt{\Gamma
_1^R\Gamma _2^R}e^{i\frac \Phi 2}\widetilde{G}_{12}^r(\varepsilon )%
\widetilde{G}_{14}^{r*}(\varepsilon )  \nonumber \\
&&-\Gamma _2^Re^{i\frac \Phi 2}\frac \Delta {\left| \varepsilon \right| }%
\widetilde{G}_{13}^r(\varepsilon )\widetilde{G}_{14}^{r*}(\varepsilon ))-%
\sqrt{\Gamma _1^R\Gamma _2^R}\frac \Delta {\left| \varepsilon \right| }%
\widetilde{G}_{11}^r(\varepsilon )\widetilde{G}_{14}^{r*}(\varepsilon ) 
\nonumber \\
&&-\sqrt{\Gamma _1^R\Gamma _2^R}\frac \Delta {\left| \varepsilon \right| }%
\widetilde{G}_{12}^r(\varepsilon )\widetilde{G}_{13}^{r*}(\varepsilon )]\}, 
\nonumber
\end{eqnarray}
\begin{eqnarray}
G_{13}^{<}(t,t) &=&i\int \frac{d\varepsilon }{2\pi }[f_L(\varepsilon
+eV)\Gamma _1^L\widetilde{G}_{11}^r(\varepsilon )\widetilde{G}%
_{13}^a(\varepsilon )+f_L(\varepsilon +eV)\sqrt{\Gamma _1^L\Gamma _2^L}e^{-i%
\frac \Phi 2}\widetilde{G}_{13}^r(\varepsilon )\widetilde{G}%
_{13}^a(\varepsilon )  \nonumber \\
&&+f_L(\varepsilon -eV)\Gamma _1^L\widetilde{G}_{12}^r(\varepsilon )%
\widetilde{G}_{23}^a(\varepsilon )+f_L(\varepsilon -eV)\sqrt{\Gamma
_1^L\Gamma _2^L}e^{i\frac \Phi 2}\widetilde{G}_{14}^r(\varepsilon )%
\widetilde{G}_{23}^a(\varepsilon )  \nonumber \\
&&+f_L(\varepsilon +eV)\sqrt{\Gamma _1^L\Gamma _2^L}e^{i\frac \Phi 2}%
\widetilde{G}_{11}^r(\varepsilon )\widetilde{G}_{33}^a(\varepsilon
)+f_L(\varepsilon +eV)\Gamma _2^L\widetilde{G}_{13}^r(\varepsilon )%
\widetilde{G}_{33}^a(\varepsilon )  \nonumber \\
&&+f_L(\varepsilon -eV)\sqrt{\Gamma _1^L\Gamma _2^L}e^{-i\frac \Phi 2}%
\widetilde{G}_{12}^r(\varepsilon )\widetilde{G}_{43}^a(\varepsilon
)+f_L(\varepsilon -eV)\Gamma _2^L\widetilde{G}_{14}^r(\varepsilon )%
\widetilde{G}_{43}^a(\varepsilon )]  \nonumber \\
&&+i\int \frac{d\varepsilon }{2\pi }f_R(\varepsilon )\widetilde{\rho }%
_R(\varepsilon )\{[\Gamma _1^R\widetilde{G}_{11}^r(\varepsilon )-\Gamma
_1^Re^{i\frac \Phi 2}\frac \Delta {\left| \varepsilon \right| }\widetilde{G}%
_{12}^r(\varepsilon )+\sqrt{\Gamma _1^R\Gamma _2^R}e^{i\frac \Phi 2}%
\widetilde{G}_{13}^r(\varepsilon )  \nonumber \\
&&-\sqrt{\Gamma _1^R\Gamma _2^R}\frac \Delta {\left| \varepsilon \right| }%
\widetilde{G}_{14}^r(\varepsilon )]\widetilde{G}_{13}^a(\varepsilon
)+[-\Gamma _1^Re^{-i\frac \Phi 2}\frac \Delta {\left| \varepsilon \right| }%
\widetilde{G}_{11}^r(\varepsilon )+\Gamma _1^R\widetilde{G}%
_{12}^r(\varepsilon )  \eqnum{B2} \\
&&-\sqrt{\Gamma _1^R\Gamma _2^R}\frac \Delta {\left| \varepsilon \right| }%
\widetilde{G}_{13}^r(\varepsilon )+\sqrt{\Gamma _1^R\Gamma _2^R}e^{-i\frac %
\Phi 2}\widetilde{G}_{14}^r(\varepsilon )]\widetilde{G}_{23}^a(\varepsilon )
\nonumber \\
&&+[\sqrt{\Gamma _1^R\Gamma _2^R}e^{-i\frac \Phi 2}\widetilde{G}%
_{11}^r(\varepsilon )-\sqrt{\Gamma _1^R\Gamma _2^R}\frac \Delta {\left|
\varepsilon \right| }\widetilde{G}_{12}^r(\varepsilon )+\Gamma _2^R%
\widetilde{G}_{13}^r(\varepsilon )  \nonumber \\
&&-\Gamma _2^Re^{-i\frac \Phi 2}\frac \Delta {\left| \varepsilon \right| }%
\widetilde{G}_{14}^r(\varepsilon )]\widetilde{G}_{33}^a(\varepsilon )+[-%
\sqrt{\Gamma _1^R\Gamma _2^R}\frac \Delta {\left| \varepsilon \right| }%
\widetilde{G}_{11}^r(\varepsilon )  \nonumber \\
&&+\sqrt{\Gamma _1^R\Gamma _2^R}e^{i\frac \Phi 2}\widetilde{G}%
_{12}^r(\varepsilon )-\Gamma _2^Re^{i\frac \Phi 2}\frac \Delta {\left|
\varepsilon \right| }\widetilde{G}_{13}^r(\varepsilon )+\Gamma _2^R%
\widetilde{G}_{14}^r(\varepsilon )]\widetilde{G}_{43}^a(\varepsilon )\}, 
\nonumber
\end{eqnarray}

\begin{eqnarray}
G_{31}^{<}(t,t) &=&i\int \frac{d\varepsilon }{2\pi }[f_L(\varepsilon
+eV)\Gamma _1^L\widetilde{G}_{31}^r(\varepsilon )\widetilde{G}%
_{11}^a(\varepsilon )+f_L(\varepsilon +eV)\sqrt{\Gamma _1^L\Gamma _2^L}e^{-i%
\frac \Phi 2}\widetilde{G}_{33}^r(\varepsilon )\widetilde{G}%
_{11}^a(\varepsilon )  \nonumber \\
&&+f_L(\varepsilon -eV)\Gamma _1^L\widetilde{G}_{32}^r(\varepsilon )%
\widetilde{G}_{21}^a(\varepsilon )+f_L(\varepsilon -eV)\sqrt{\Gamma
_1^L\Gamma _2^L}e^{i\frac \Phi 2}\widetilde{G}_{34}^r(\varepsilon )%
\widetilde{G}_{21}^a(\varepsilon )  \nonumber \\
&&+f_L(\varepsilon +eV)\sqrt{\Gamma _1^L\Gamma _2^L}e^{i\frac \Phi 2}%
\widetilde{G}_{31}^r(\varepsilon )\widetilde{G}_{31}^a(\varepsilon
)+f_L(\varepsilon +eV)\Gamma _2^L\widetilde{G}_{33}^r(\varepsilon )%
\widetilde{G}_{31}^a(\varepsilon )  \nonumber \\
&&+f_L(\varepsilon -eV)\sqrt{\Gamma _1^L\Gamma _2^L}e^{-i\frac \Phi 2}%
\widetilde{G}_{32}^r(\varepsilon )\widetilde{G}_{41}^a(\varepsilon
)+f_L(\varepsilon -eV)\Gamma _2^L\widetilde{G}_{34}^r(\varepsilon )%
\widetilde{G}_{41}^a(\varepsilon )]  \nonumber \\
&&+i\int \frac{d\varepsilon }{2\pi }f_R(\varepsilon )\widetilde{\rho }%
_R(\varepsilon )\{[\Gamma _1^R\widetilde{G}_{31}^r(\varepsilon )-\Gamma
_1^Re^{i\frac \Phi 2}\frac \Delta {\left| \varepsilon \right| }\widetilde{G}%
_{32}^r(\varepsilon )+\sqrt{\Gamma _1^R\Gamma _2^R}e^{i\frac \Phi 2}%
\widetilde{G}_{33}^r(\varepsilon )  \nonumber \\
&&-\sqrt{\Gamma _1^R\Gamma _2^R}\frac \Delta {\left| \varepsilon \right| }%
\widetilde{G}_{34}^r(\varepsilon ))]\widetilde{G}_{11}^a(\varepsilon
)+[-\Gamma _1^Re^{-i\frac \Phi 2}\frac \Delta {\left| \varepsilon \right| }%
\widetilde{G}_{31}^r(\varepsilon )+\Gamma _1^R\widetilde{G}%
_{32}^r(\varepsilon )  \eqnum{B3} \\
&&-\sqrt{\Gamma _1^R\Gamma _2^R}\frac \Delta {\left| \varepsilon \right| }%
\widetilde{G}_{33}(\varepsilon )+\sqrt{\Gamma _1^R\Gamma _2^R}e^{-i\frac \Phi
2}\widetilde{G}_{34}^r(\varepsilon )]\widetilde{G}_{21}^a(\varepsilon ) 
\nonumber \\
&&+[\sqrt{\Gamma _1^R\Gamma _2^R}e^{-i\frac \Phi 2}\widetilde{G}%
_{31}^r(\varepsilon )-\sqrt{\Gamma _1^R\Gamma _2^R}\frac \Delta {\left|
\varepsilon \right| }\widetilde{G}_{32}^r(\varepsilon )+\Gamma _2^R%
\widetilde{G}_{33}^r(\varepsilon )  \nonumber \\
&&-\Gamma _2^Re^{-i\frac \Phi 2}\frac \Delta {\left| \varepsilon \right| }%
\widetilde{G}_{34}^r(\varepsilon )]\widetilde{G}_{31}^a(\varepsilon )+[-%
\sqrt{\Gamma _1^R\Gamma _2^R}\frac \Delta {\left| \varepsilon \right| }%
\widetilde{G}_{31}^r(\varepsilon )  \nonumber \\
&&+\sqrt{\Gamma _1^R\Gamma _2^R}e^{i\frac \Phi 2}\widetilde{G}%
_{32}^r(\varepsilon )-\Gamma _2^Re^{i\frac \Phi 2}\frac \Delta {\left|
\varepsilon \right| }\widetilde{G}_{33}^r(\varepsilon )+\Gamma _2^R%
\widetilde{G}_{34}^r(\varepsilon )]\widetilde{G}_{41}^a(\varepsilon )\}, 
\nonumber
\end{eqnarray}
\begin{eqnarray}
G_{33}^{<}(t,t) &=&i\int \frac{d\varepsilon }{2\pi }\{f_L(\varepsilon
+eV)\Gamma _1^L\left| \widetilde{G}_{31}^r(\varepsilon )\right|
^2+f_L(\varepsilon -eV)\Gamma _1^L\left| \widetilde{G}_{32}^r(\varepsilon
)\right| ^2  \nonumber \\
&&+f_L(\varepsilon +eV)\Gamma _2^L\left| \widetilde{G}_{33}^r(\varepsilon
)\right| ^2+f_L(\varepsilon -eV)\Gamma _2^L\left| \widetilde{G}%
_{34}^r(\varepsilon )\right| ^2  \nonumber \\
&&+2\sqrt{\Gamma _1^L\Gamma _2^L}%
\mathop{\rm Re}
[f_L(\varepsilon +eV)e^{i\frac \Phi 2}\widetilde{G}_{31}^r(\varepsilon )%
\widetilde{G}_{33}^{r*}(\varepsilon )  \nonumber \\
&&+f_L(\varepsilon -eV)e^{-i\frac \Phi 2}\widetilde{G}_{32}^r(\varepsilon )%
\widetilde{G}_{34}^{r*}(\varepsilon )]\}  \nonumber \\
&&+i\int \frac{d\varepsilon }{2\pi }f_R(\varepsilon )\widetilde{\rho }%
_R(\varepsilon )\{\Gamma _1^R\left| \widetilde{G}_{31}^r(\varepsilon
)\right| ^2+\Gamma _1^R\left| \widetilde{G}_{32}^r(\varepsilon )\right| ^2 
\eqnum{B4} \\
&&+\Gamma _2^R\left| \widetilde{G}_{33}^r(\varepsilon )\right| ^2+\Gamma
_2^R\left| \widetilde{G}_{34}^r(\varepsilon )\right| ^2  \nonumber \\
&&+2%
\mathop{\rm Re}
[-\Gamma _1^Re^{i\frac \Phi 2}\frac \Delta {\left| \varepsilon \right| }%
\widetilde{G}_{32}^r(\varepsilon )\widetilde{G}_{31}^{r*}(\varepsilon )+%
\sqrt{\Gamma _1^R\Gamma _2^R}e^{i\frac \Phi 2}\widetilde{G}%
_{33}^r(\varepsilon )\widetilde{G}_{31}^{r*}(\varepsilon )  \nonumber \\
&&+\sqrt{\Gamma _1^R\Gamma _2^R}e^{i\frac \Phi 2}\widetilde{G}%
_{32}^r(\varepsilon )\widetilde{G}_{34}^{r*}(\varepsilon )-\Gamma _2^Re^{i%
\frac \Phi 2}\frac \Delta {\left| \varepsilon \right| }\widetilde{G}%
_{33}^r(\varepsilon )\widetilde{G}_{34}^{r*}(\varepsilon )  \nonumber \\
&&-\sqrt{\Gamma _1^R\Gamma _2^R}\frac \Delta {\left| \varepsilon \right| }%
\widetilde{G}_{32}^r(\varepsilon )\widetilde{G}_{33}^{r*}(\varepsilon )-%
\sqrt{\Gamma _1^R\Gamma _2^R}\frac \Delta {\left| \varepsilon \right| }%
\widetilde{G}_{31}^r(\varepsilon )\widetilde{G}_{34}^{r*}(\varepsilon )]\}. 
\nonumber
\end{eqnarray}

The results here show that the above matrix elements of $G^{<}(t,t)$ do not
depend on time $t$, because for the N-QD-S hybrid system, the current should
be independent of time for the dc bias \cite{15}.

FIGURE CAPTION

FIG. 1: Schematic diagram for the double AB interfermeter connected with
normal metal and the superconductor leads, respectively.

FIG. 2: The probability of the Andreev reflection $T_{AR}$ vs $\varepsilon $
in the unit of the energy gap $\Delta $ for (a) the symmetry case( $\Gamma
_{1}^{L}=\Gamma _{1}^{R}=$\ $\Gamma _{2}^{L}=\Gamma _{2}^{R}=0.02$) and
asymmetry case (b) $\Gamma _{1}^{L}=$\ $\Gamma _{2}^{L}=0.08$,$\Gamma
_{1}^{R}=\Gamma _{2}^{R}=0.02$, (c) $\Gamma _{1}^{L}=\Gamma _{1}^{R}=0.08$, $%
\Gamma _{2}^{L}=\Gamma _{2}^{R}=0.02$. The solid, dotted, dot-dashed, and
dashed lines correspond to $\Omega $=0, 0.02, 0.04, and 0.08, respectively.

FIG. 3: The magnetic flux $\Phi $ dependence of the Andreev reflection
probability for $\Gamma _{1}^{L}=\Gamma _{1}^{R}=$ $\Gamma _{2}^{L}=\Gamma
_{2}^{R}=0.02$, $\varepsilon _{1}=\varepsilon _{2}=0$, $\Omega =0.2$, and $%
\Phi _{12}=0$, solid line: $\Phi =\pi /3$, dotted line: $\Phi =2\pi /3$.

FIG. 4: The periodic oscillation of the Andreev reflection probability $%
T_{AR}$ with the magnetic flux $\Phi $ ( $\Gamma _1^L=\Gamma _1^R=\Gamma
_2^L=\Gamma _2^R=0.02$, $\varepsilon _1=-0.1$, $\varepsilon _2=0.1$), (a) $%
\alpha _1/\alpha _2=1$. Solid line: $\Omega =0$, $\varepsilon =0.1$, dotted
line: $\Omega =0.05$, $\varepsilon =0.11$, and dashed line: $\Omega =0.1$, $%
\varepsilon =0.14$. (b) $\Omega =0.05$, $\varepsilon =0.11$. Solid line: $%
\alpha _1/\alpha _2=2$, dotted line: $\alpha _1/\alpha _2=3$, and $4$ for
dashed line.

FIG. 5: Paths and corresponding phase shift. (a) for the incident electron
from left lead to superconductor (b) for the reflecting hole from
superconductor to left lead.

FIG. 6: Andreev reflection current $J_{A}$ vs gate voltage $V_{g}$ ( $%
\varepsilon _{1}=0.1$, $\varepsilon _{2}=0.3$, $\Gamma _{1}^{L}=\Gamma
_{1}^{R}=$ $\Gamma _{2}^{L}=\Gamma _{2}^{R}=0.02$, and $\Phi =\Phi _{12}=0$)
for bias voltage $V=0.4$, $\Omega =0$ (solid line), $\Omega =0.05$ (dotted
line), and $\Omega =0.1$ (dashed line).

FIG. 7: Andreev reflection current vs gate voltage with the fixed total
magnetic flux that $\Phi =\pi /2$ for $\Omega =0.05$. $\Phi _{12}=0$ (solid
line), $\pi /2$ (dotted line), $3\pi /4$ (dot-dashed line), $\pi $ (dashed
line).

FIG. 8: Andreev reflection current vs gate voltage for $\Phi =3\pi /2$. $%
\Phi _{12}=0$ (solid line), $\pi /4$ (short-dashed line), $\pi /2$ (dotted
line), $\pi $ (long-dashed line).

FIG. 9: Andreev reflection current vs gate voltage with $\Phi _{12}=\pi /5$
for $\Phi =0$ (solid line), $2\pi /3$ (dotted line), $5\pi /3$ (dashed line).

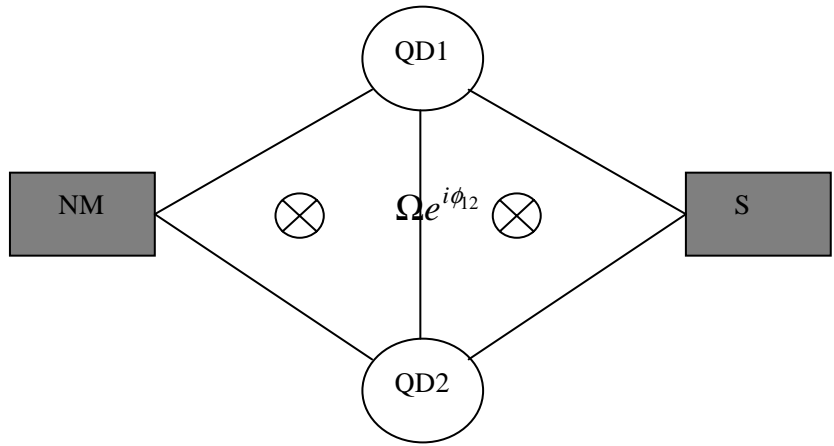
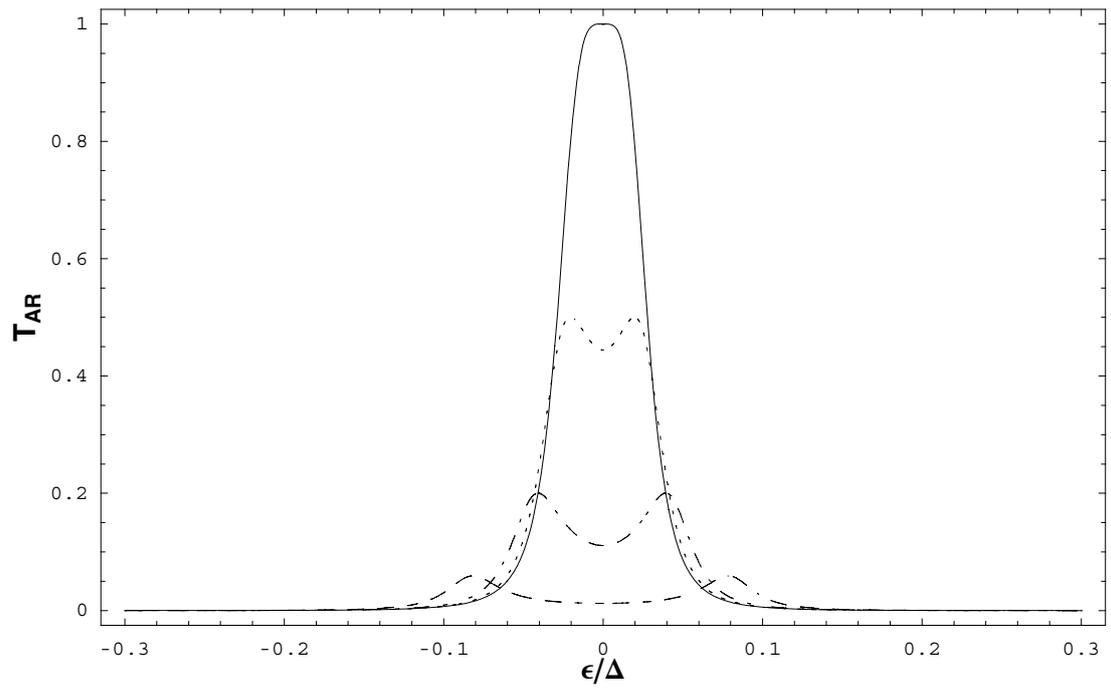
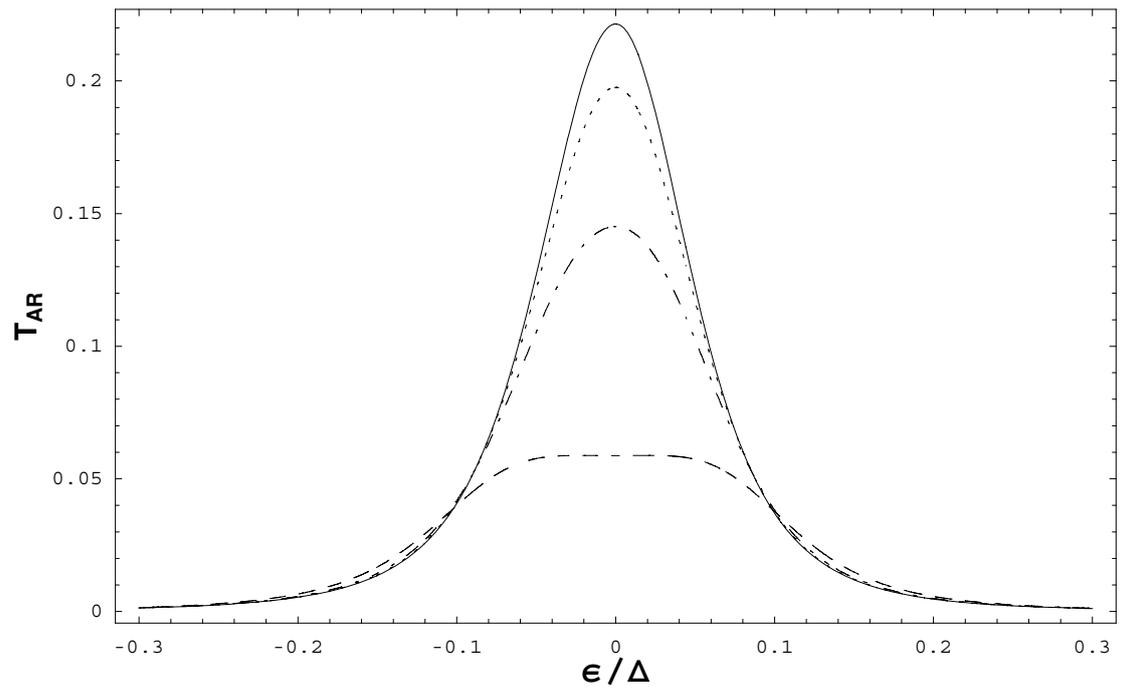
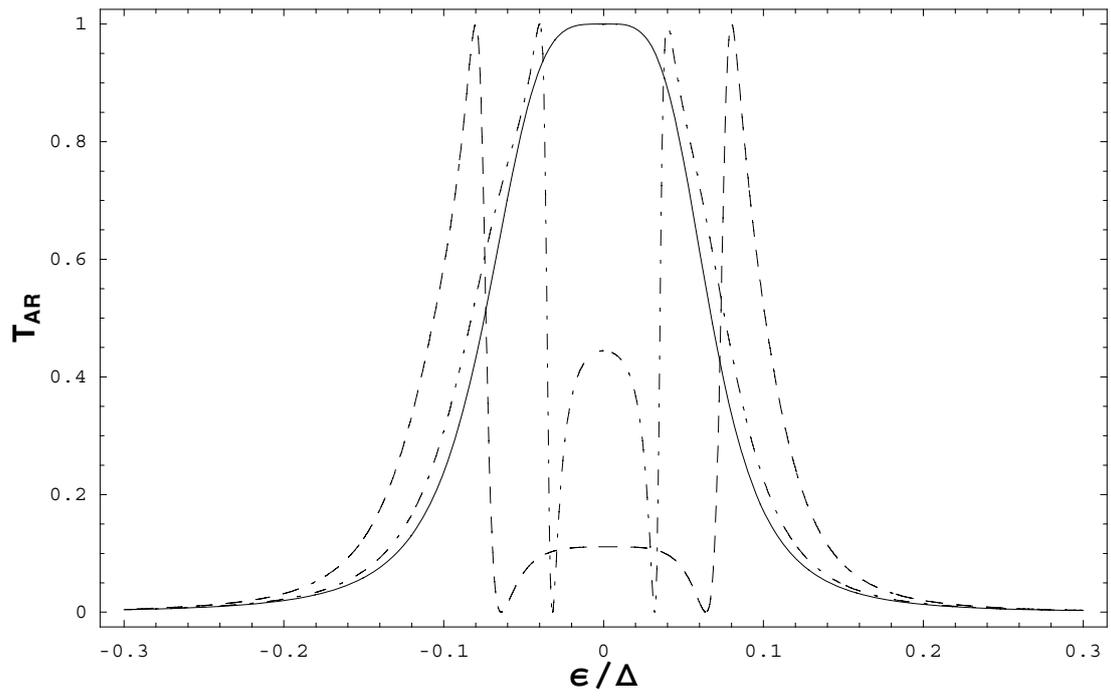
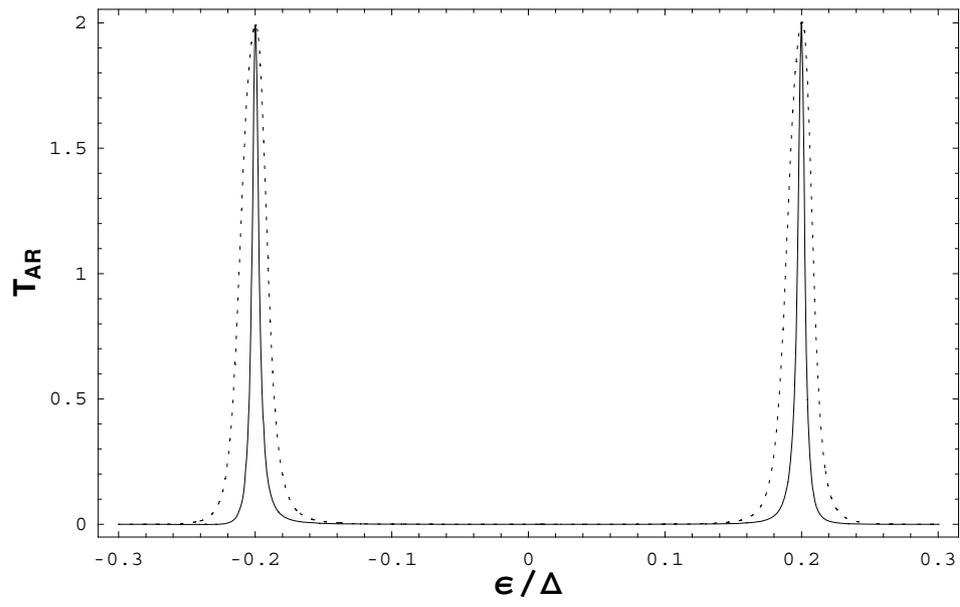
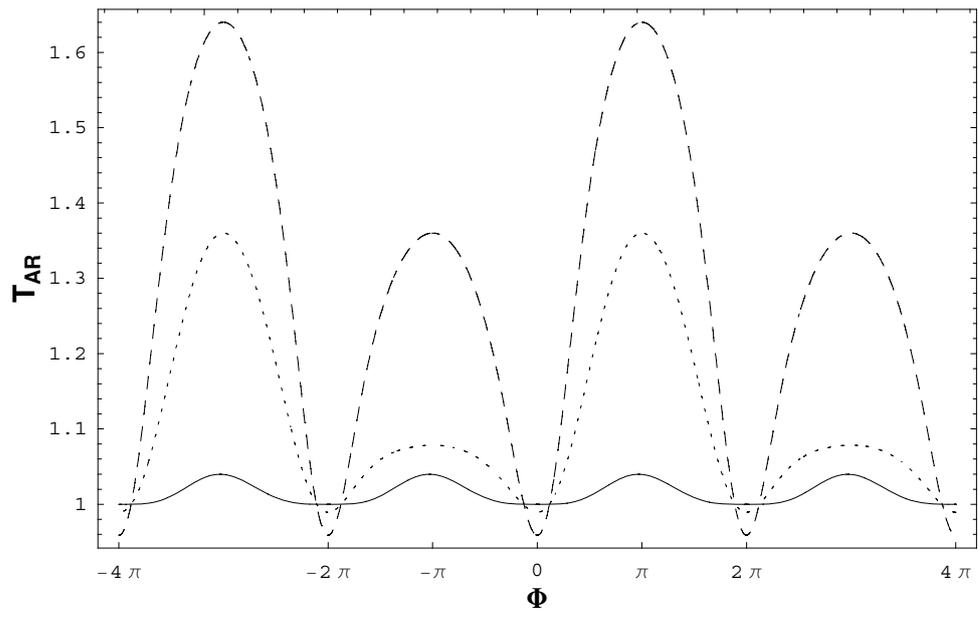
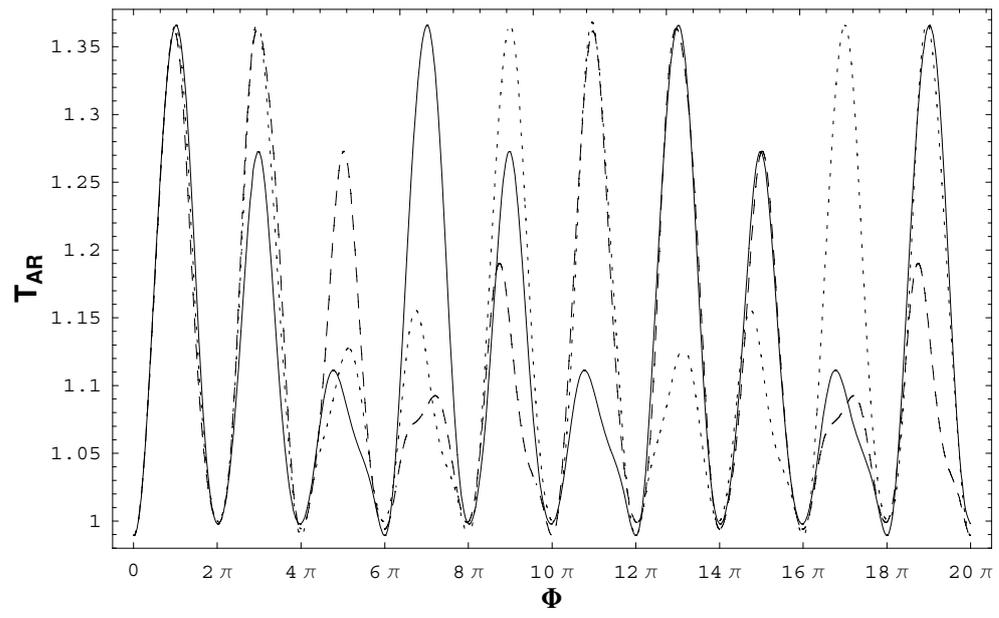
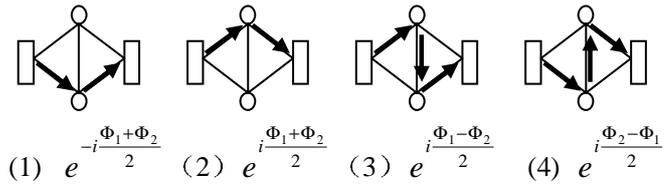
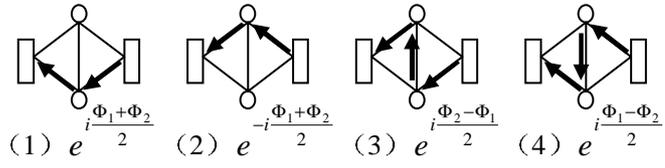
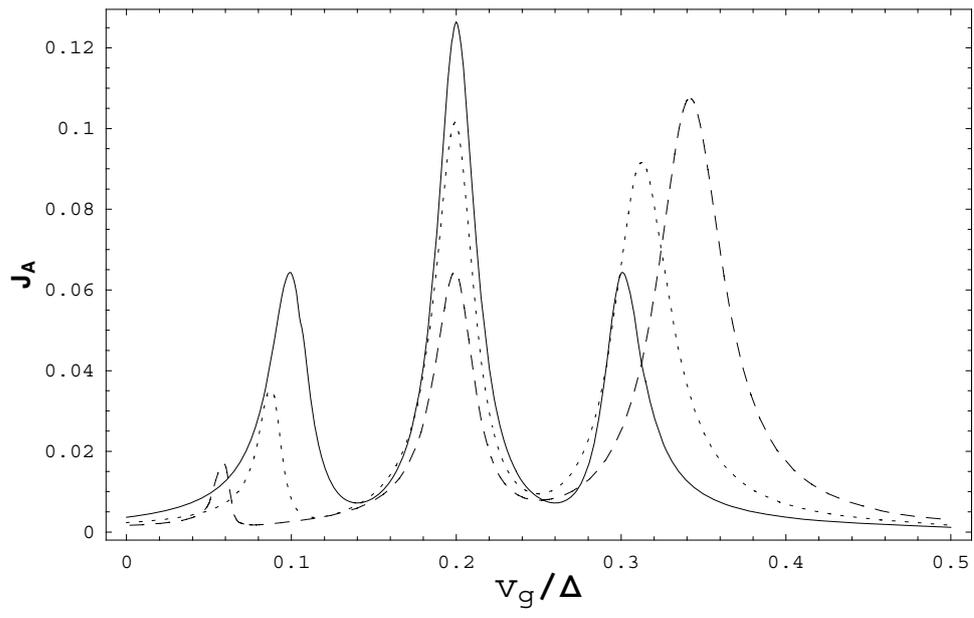
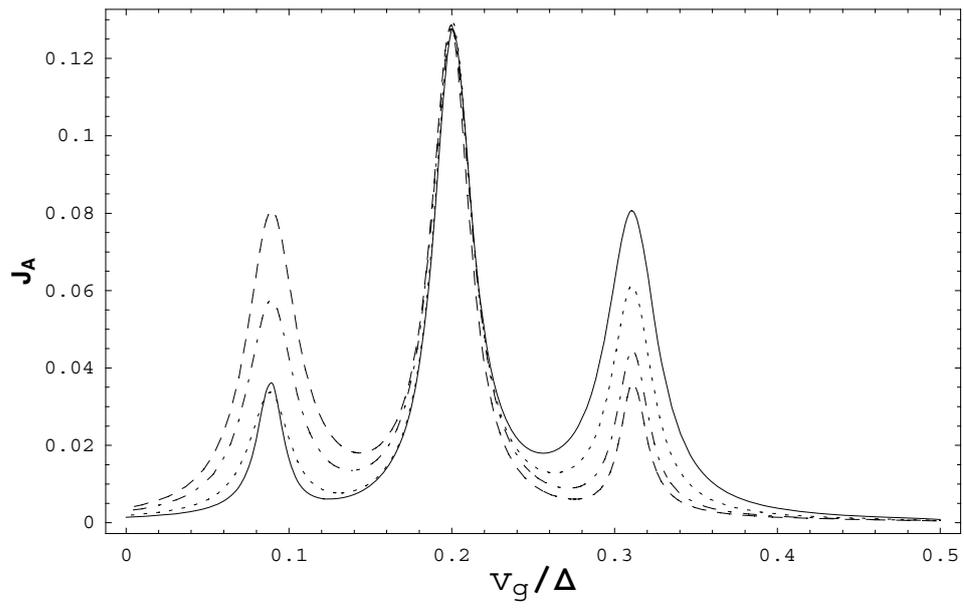
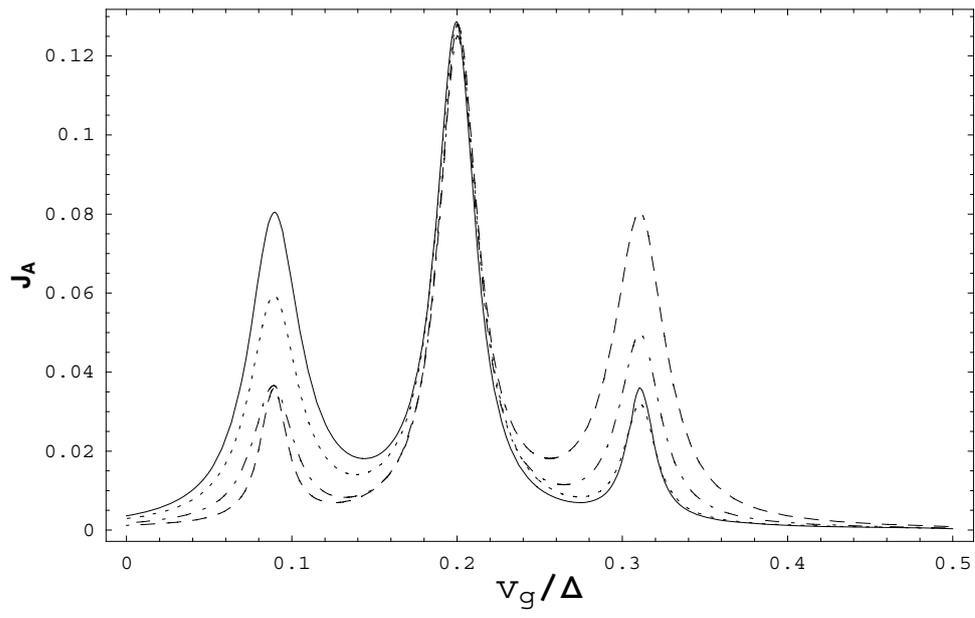
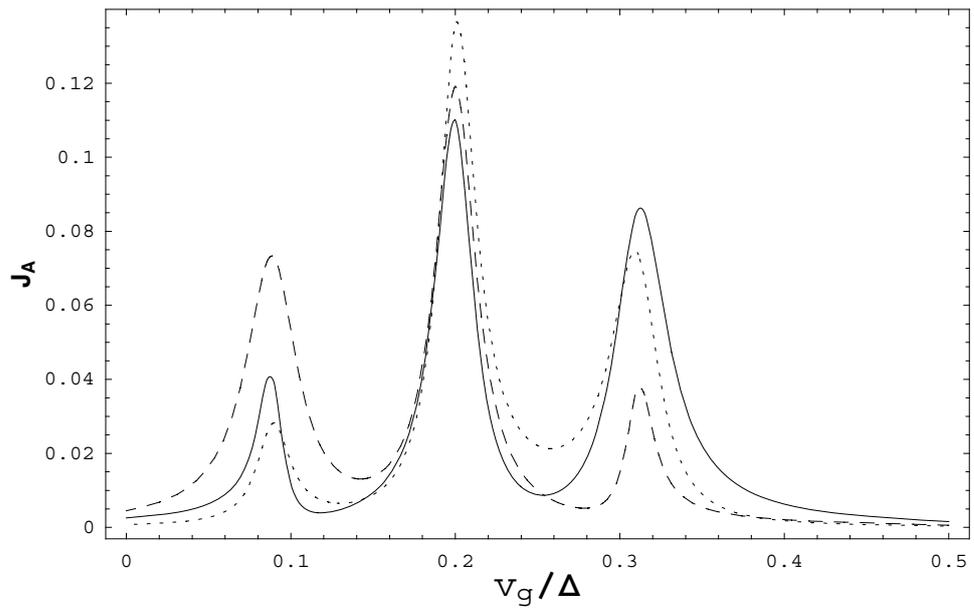
\end{document}